\documentclass[10pt, conference, compsocconf]{IEEEtran}
\usepackage{mathptmx} %
\let\labelindent\relax
\usepackage{enumitem,kantlipsum}

\usepackage{amsmath}
\usepackage{tcolorbox}
\usepackage{fancyhdr}
\usepackage[normalem]{ulem}
\usepackage[hyphens]{url}
\usepackage[sort,nocompress]{cite}
\usepackage[final]{microtype}
\usepackage{flushend}
\usepackage[bookmarks=true,breaklinks=true,letterpaper=true,colorlinks,citecolor=blue,linkcolor=blue,urlcolor=blue]{hyperref}

\title{Guidelines for Submission to MICRO 2022} 

\usepackage{cite}
\usepackage{amsmath,amssymb,amsfonts}
\usepackage{algorithmicx}
\usepackage{graphicx}
\usepackage{textcomp}
\usepackage{xcolor}
\usepackage[hyphens]{url}
\usepackage{xspace}
\usepackage{setspace}

\usepackage[noend]{algpseudocode}

\usepackage{flushend}
\usepackage{color}
\usepackage{makecell}
\usepackage{adjustbox}
\usepackage{multirow}
\usepackage{booktabs}

\usepackage{lipsum}
\usepackage{titlesec}
\def\BibTeX{{\rm B\kern-.05em{\sc i\kern-.025em b}\kern-.08em
    T\kern-.1667em\lower.7ex\hbox{E}\kern-.125emX}}

\newcommand{\para}[1]{\noindent\textbf{{#1}\xspace}}
\newcommand{\scheme}{\textsc{{E}vax}}

\newcommand{\ignore}[1]{}

\newenvironment{CompactItemize}%
{\begin{list}{$\bullet$}%
  {\leftmargin=\parindent \itemsep=2pt \topsep=2pt
    \parsep=0pt \partopsep=0pt}}%
{\end{list}}

\usepackage{todonotes}
\setuptodonotes{inline}

\usepackage{enumitem}
 \usepackage{subfigure}

\usepackage{pdfpages}
\makeatletter
\newcommand{\linebreakand}{%
  \end{@IEEEauthorhalign}
  \hfill\mbox{}\par
  \mbox{}\hfill\begin{@IEEEauthorhalign}
}
\makeatother

\definecolor{ao}{rgb}{0,0,0}

\ifCLASSINFOpdf
\else
\fi
\begin{document}

\titlespacing\subsection{2pt}{1.5pt}{1.5pt}
\title{%
~\scheme{}: Towards a Practical, Pro-active \& Adaptive Architecture for High Performance \& Security}

\author{\IEEEauthorblockN{Samira Mirbagher Ajorpaz}
\IEEEauthorblockA{
North Carolina State University,
USA\\
smirbag@ncsu.edu}\and
\IEEEauthorblockN{Daniel Moghimi}
\IEEEauthorblockA{
University of California San Diego, USA\\
danielm@ucsd.edu}\and
\IEEEauthorblockN{Jeffrey Neal Collins}
\IEEEauthorblockA{Independent Author, USA\\
collinsje91@gmail.com}
\linebreakand %
\IEEEauthorblockN{Gilles Pokam}
\IEEEauthorblockA{Intel Labs, USA\\
gilles.a.pokam@intel.com}\and
\IEEEauthorblockN{Nael Abu-Ghazaleh}
\IEEEauthorblockA{
{University of California Riverside,}
USA \\
naelag@ucr.edu}\and
\IEEEauthorblockN{Dean Tullsen}
\IEEEauthorblockA{
University of California San Diego,
USA\\
tullsen@cs.ucsd.edu }
}

\maketitle
\begingroup
\renewcommand{\thefootnote}{}
\footnotetext{Author manuscript of the 2022 55th IEEE/ACM International Symposium on Microarchitecture (MICRO), pp. 1218--1236 (2022). DOI: \url{https://doi.org/10.1109/MICRO56248.2022.00085}. \textcopyright\ 2022 IEEE. Personal use of this material is permitted. Permission from IEEE must be obtained for all other uses, in any current or future media, including reprinting/republishing this material for advertising or promotional purposes, creating new collective works, for resale or redistribution to servers or lists, or reuse of any copyrighted component of this work in other works.}
\endgroup

\IEEEpeerreviewmaketitle

 \begin{abstract}
This paper provides an end-to-end solution to defend against known microarchitectural attacks such as speculative execution attacks, fault-injection attacks, covert and side channel attacks, and unknown or evasive versions of these attacks.
  Current defenses are attack specific and can have unacceptably high performance overhead. We propose an approach that reduces the  overhead of state-of-art defenses by over 95\%, by applying defenses only when attacks are detected. 
  Many current proposed mitigations are not practical for deployment; for example, InvisiSpec has 27\% overhead and Fencing has 74\% overhead while protecting against only Spectre attacks. Other mitigations carry similar performance penalties. 
  We reduce the overhead for InvisiSpec to 1.26\% and for Fencing to 3.45\% offering {\em performance and security} for not only spectre attacks but other known transient attacks as well, including the dangerous class of LVI and Rowhammer attacks, as well as covering a large set of future evasive and  zero-day attacks.

  Critical to our approach is an accurate detector that is not fooled by evasive attacks and that can generalize to novel zero-day attacks.  We use a novel Generative framework, Evasion Vaccination (~\scheme{}) for training ML models and engineering new security-centric performance counters. \scheme{} significantly increases sensitivity to detect and classify attacks in time for mitigation to be deployed with low false positives (4 FPs in every 1M instructions in our experiments). Such performance enables efficient and timely mitigations, enabling the processor to automatically switch between performance and security as needed. 
\end{abstract}

\begin{IEEEkeywords}
Hardware Security,
Side channel, Generative Adversarial Networks, Automatic Attack Sample Generation, Adversarial Machine Learning Attacks, Automated Hardware Performance Counter Engineering, Microarchitectural Attack Detection, Linearized Neural Network, ML Interpretability, Zero Day Attack Defense. 

\end{IEEEkeywords}

\section{Introduction}

The cost of Global Cybercrime is growing at a rate of 15\% per year.
The average cost of a data breach in 2020 was \$3.86 million with malicious Attacks being responsible for 52\% of data breaches
~\cite{IBMData2020}.
More than 100 Zettabytes of data is distributed across clouds ~\cite{IBMData2020}. %
In 2018 with the disclosure of {\em transient} (e.g., Meltdown~\cite{meltdown} and Spectre~\cite{kocher2019spectre}) attacks, we learned that most processors currently in use are insecure.

Microarchitectural attacks can be extremely dangerous.  For example, transient attacks can compromise all secret data on a system: %
they can leak  arbitrary memory locations from  kernel or victim memory. Other attacks such as RowHammer go beyond data confidentiality and break the integrity of programs by inducing controlled memory errors in DRAM cells~\cite{rowhammer}\cite{gruss2016rowhammer}.
Most microarchitectural attacks %
can not be fully mitigated in software.  For example, studies have shown that  all currently deployed Rowhammer mitigations are still vulnerable to attack~\cite{frigo2020trrespass}\cite{cojocar2019exploiting}. 

Microarchitectural attacks continue to emerge at a breathtaking pace.  Even worse, newly developed automated attack generation tools %
can discover new vulnerabilities, %
challenging existing defenses~\cite{gras2020absynthe}\cite{moghimi2020medusa}\cite{TRRespass}\cite{weber2021osiris}.
Many of these generated attacks have no mitigation~\cite{weber2021osiris}, {\em e.g.} FlushConflict breaks even the newest Intel Ice Lake and Comet Lake micro-architectures which are hardened against Spectre and Meltdown.

Transient attacks, to pick an example class of microarchitectural attacks, target the foundations of hardware optimization. Mitigating them fully leads to a significant trade off between performance and security; e.g., 
21\% - 72\% performance overhead for the state-of-the-art HW mitigation for Spectre-type attacks (InvisiSpec) and 74\% - 204\% for Fencing~\cite{yan2018invisispec}. 
The overhead for mitigating LVI attack is over 900\% ~\cite{vanbulck2020lvi}.  
The number of known security vulnerabilities caused by speculative execution has increased sharply in recent years~\cite{canella2018systematic}. %
New variants of these \emph{transient execution attacks}   undermine system security boundaries~\cite{canella2018systematic,canella2019fallout,schwarz2019zombieload,ridl,van2018foreshadow,moghimi2020medusa,vanbulck2020lvi}.  Other variations exploit other microarchitectural optimizations and their shared hardware components ~\cite{osvik2006cache,aciiccmez2007yet,ren2021see, gras2018translation, evtyushkin2018branchscope, evtyushkin2016jump, aldaya2019port, bhattacharyya2019smotherspectre, islam2019spoiler, moghimi2019memjam, pessl2016drama, paccagnella2021lord, schwarz2018keydrown, irazoqui2015s, liu2015last}. 
 Defenses proposed to date~\cite{yan2018invisispec,koruyeh2019speccfi,kiriansky2018dawg,domnitser2012non,saileshwar2019cleanupspec} only target specific attack variants or protect individual microarchitectural components. Complete protection requires %
multiple mitigations that are active concurrently,  imposing additive performance overhead on the processor. This paper proposes an ML-assisted architecture that can detect attacks and apply targeted mitigations only as needed significantly reducing their performance overhead.

Currently, the average time to identify a data breach is 280 days~\cite{IBMData2020}. The average cost savings of containing a breach in less than 200 days vs.  more than 200 days is \$1.12 million ~\cite{IBMData2020}.  The propose detection architecture will %
be part of a proactive and an adaptive end-to-end solution for security closing the gap in detection and eliminating the significant cost of reactive security response. 

Detection of microarchitectural attacks poses several challenges. Attacks can have short signatures.  For example, transient attacks trigger transient execution (using  branch prediction or out-of-order execution).  During the transient window, they complete their attack and leak secret data  to the attacker. 
Access to a high number of hardware performance counters (HPCs) monitoring transient events (uncommitted instructions) is essential. Lightning fast Sampling and classification %
is essential for mitigations to be applied prior to leakage. Finally, detection can be improved by engineering new security-centric HPCs to detect new attacks.    
Software detectors %
fail because the number of events that can be monitored from software without multiplexing counters is limited (up to 4). Computation is on the critical path, limiting the sampling frequency %
and leaving software detectors vulnerable to bandwidth evasion ~\cite{Gaudiot2020, PerSpectron}. %
On the other hand, hardware detectors are promising due to significantly higher sampling frequency and computations being conducted outside the critical path ~\cite{ PerSpectron}. We show that prior work fails to detect evasive attacks and many novel attacks, due to insufficient training data and dependence on current HPCs 
which were designed for tuning performance. We need to develop new security-centric HPCs that expose features that promote accurate and rapid detection. 

Perceptron-based detectors have been shown to provide classification outputs within the transient window~\cite{PerSpectron} without computational delay. PerSpectron was able to negate bandwidth evasion due to its high sampling frequency on the order of 100 nsec to 1$\mu$-sec ~\cite{PerSpectron}. %
However, PerSpectron is limited in detecting zero-day attacks and is susceptible to evasive attacks generated by new {\em fuzzing technologies} and AML attacks ~\cite{Carlini2017,RHMD2017,gras2020absynthe,moghimi2020medusa,TRRespass,weber2021osiris}.

  \begin{figure}[!htbp]
    \centering
    \includegraphics[width=0.4\textwidth]{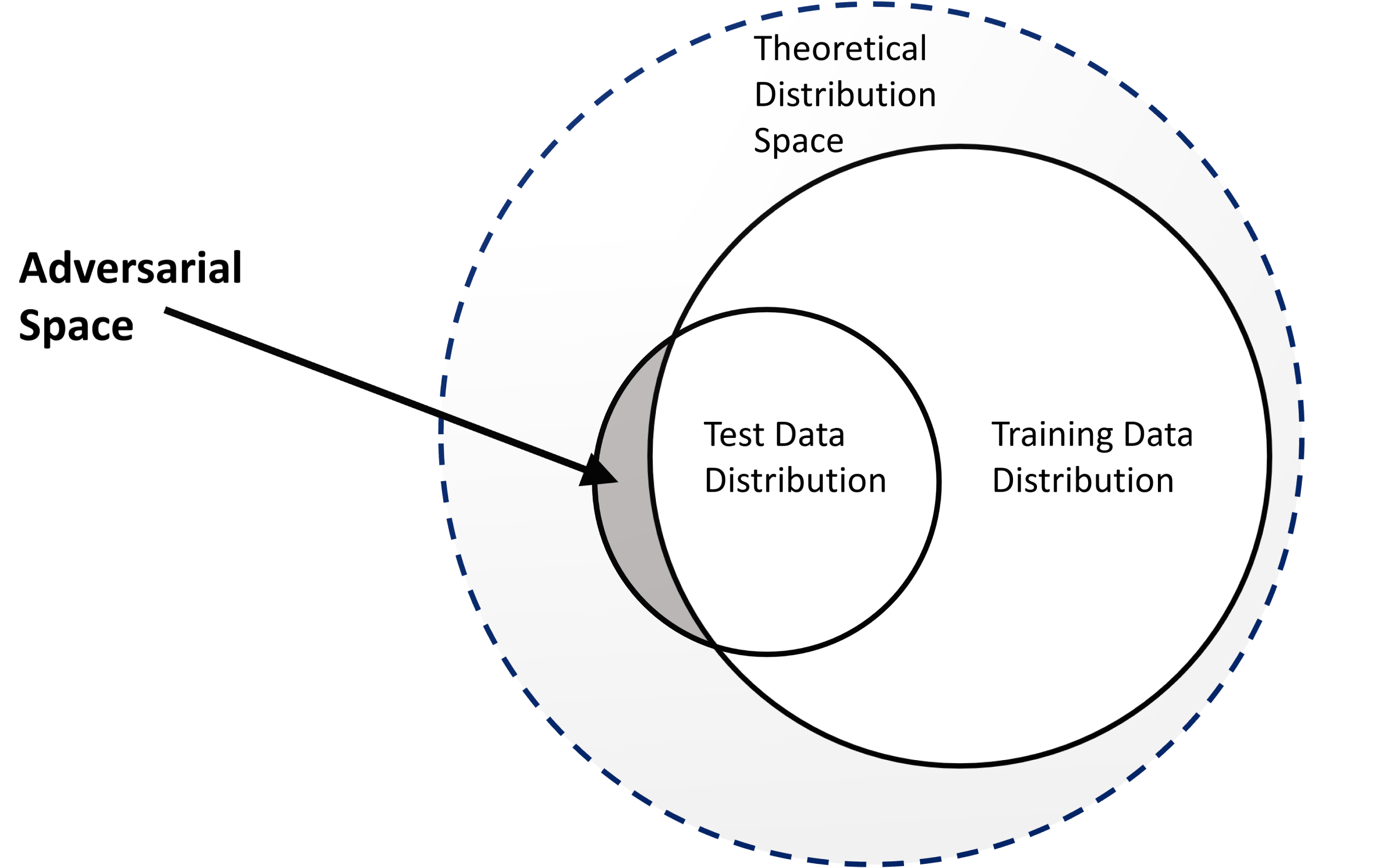}
    \caption{Root Cause of ML Vulnerability}
    \label{fig:AML}
\end{figure}

\begin{figure}[!hbp]
    \centering
    \includegraphics[width=0.4\textwidth]{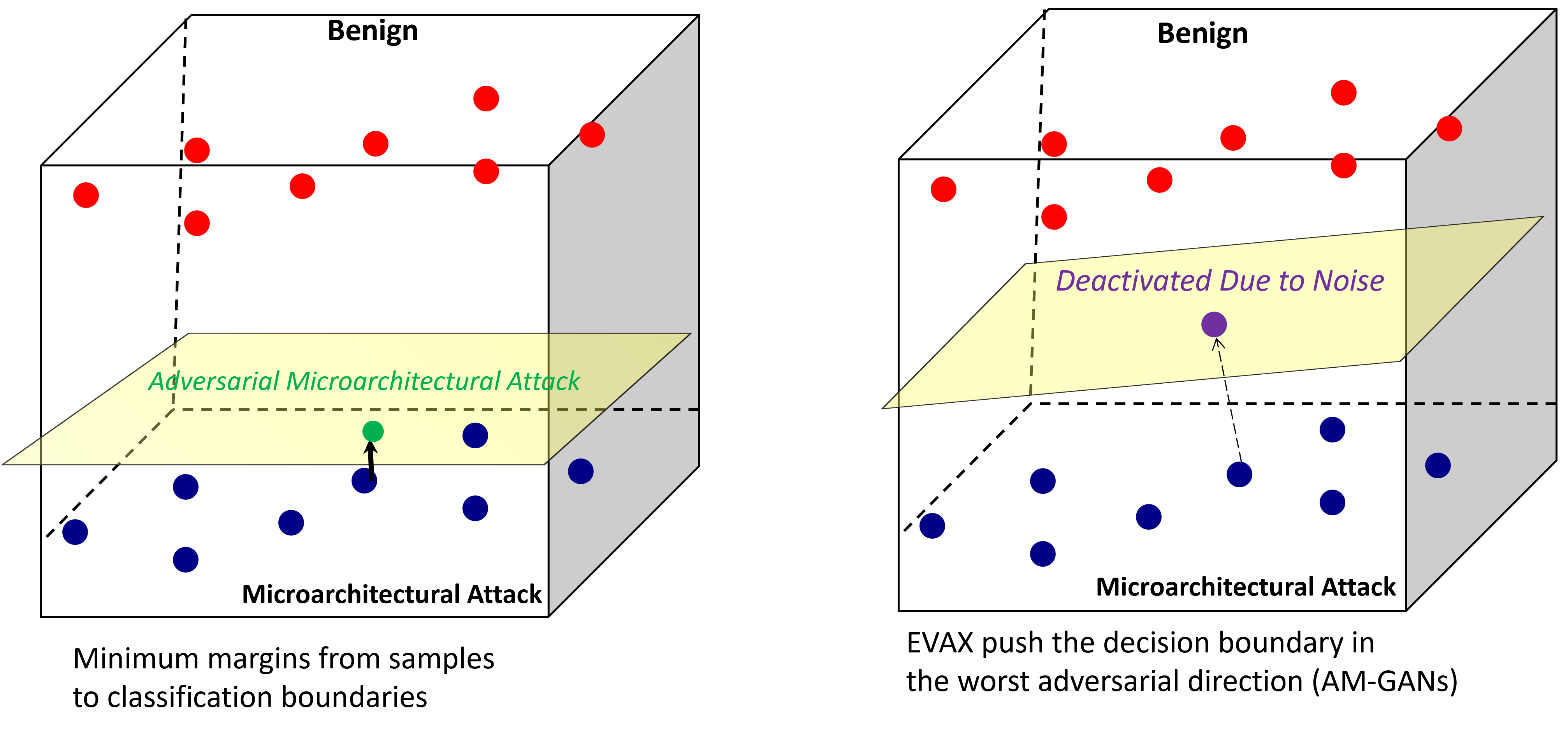}
    \caption{Large classification margin toward adversarial directions, to disable timing-sensitive attacks. }
    \label{fig:margin}
\end{figure}

Attackers exploit pockets of adversarial space between the data fitted by the learning agent and the theoretical distribution space to fool ML (see Figure~\ref{fig:AML}) .  A large class of vulnerabilities arise from the fundamental problems of imperfect learning:
{(1) Ineffectiveness of classifier models to contribute to the adversarial space; and
(2) Flawed adversarial space identification as a result of an imperfect training data set~\cite{MLSecurity2018}.
} 
We need automatic attack sample generation tools to provide a
broad range of possible permutations/variants of microarchitectural attacks to train our classifier. %
Prior works %
have proposed automatic attack code generation techniques that %
attempt to create new attacks~\cite {gras2020absynthe,moghimi2020medusa,TRRespass,weber2021osiris}. %
These transformations applied at the source or binary level, include insertion, deletion, rewriting, and other strategies, using a fuzzing-based approach. Fuzzing tools are typically used to permute a single
type of microarchitectural attacks, for example, Meltdown
or Rowhammer-type attacks. %
Attacks generated by these tools do not comprehensively explore the evasion space, missing a slew of potentialy dangerous attacks or evasion strategies. This is because code transformations ~\cite {gras2020absynthe,moghimi2020medusa,TRRespass,weber2021osiris} do %
not automatically translate to the complex feature space of the classifier.  
\para{Our solution--- Disabling Adversarial ML Attacks by Increased Classification Margin beyond the Leakage Window. } 
Evasive attacks exploit the minimum margins from samples to classification
boundaries (see figure~\ref{fig:margin}). Recognizing that microarchitectural attacks are timing sensitive-- {\em they  rely on bringing the microarchitecture to specific state during a short transient window and transferring the secret before activity of other processes and kernel destroys its footprints}, our solution is to  push the classification boundaries in
the worst {\em adversarial} directions until further attempts to evade disables the attack. We do that by retraining the classifier on a very large set of adversarial microarchitectural samples automatically generated through a comprehensive search using Generative Adversarial Networks (GANs)~\cite{goodfellow2014generative}. 

Microarchitectural attackers can only leak data for a limited time interval before the CPU eventually catches up, detects the fault, or aborts transient execution. This implies that
there is only a limited transient window in which the victim can leak data e.g., in LVI victim inadvertently
computes on the poisoned load values, and all required gadget instructions
need to complete within this window to transmit secrets. The transient
window is ultimately bounded by the size of the processor's reorder buffer (ROB). For example, our experiments show adversarial ML efforts in systems with small ROB fail to evade our detector. ROB size reflects the boundary of transient window and is a property of system that can be learned through self-supervised learning (e.g, Generative Modeling).

We introduce the Evasive Vaccination framework (~\scheme{}) which generates adversarial examples that push decision boundaries in
adversarial directions automatically. 
To identify the adversarial directions we leverage Generative Adversarial Networks (GANs). GAN contains two different networks; a Generator and a Discriminator which play an adversarial game against each other. 
In GAN, two models have opposite agendas: discrimination and generation. During each step of the training, the output of each model improves the other.
Because the training of the generator in GAN is toward
gradient descent (adversarial direction),
retraining with the generated examples that fool the discriminator, enlarges the margins of decision boundaries
to the point that code transformations needed for evasion disable the attack.
~\scheme{} can fine-tune the generation of new microarchitectural attack samples to virtually {\em infinite} granularity.

In line with the ethical guidelines of Google’s SafeSide project~\cite{GoogleSafe}, 
~\scheme{} aims to create attack samples that are useful for training and analysis but cannot be weaponized by attackers. ~\scheme{} Generated samples are vectors of microarchitectural counter values. With no direct mapping between code transformations and feature values an attacker cannot reverse engineer attack code %
from our attack samples. In contrast, the traditional automatic attack generation frameworks~\cite {gras2020absynthe,moghimi2020medusa,TRRespass,weber2021osiris} %
can provide attackers with  powerful tools to compromise even a hardened system. %

  We also showed that when the counters representing the inner-nodes of a Generator deep NN in GANs are combined and moved to the input-layer, the accuracy of the linear NN (fast and HW friendly) becomes equivalent to a deeper NN. 
Interestingly, we concurrently observed and showed empirically that the same property proved by the recent GoogleMind study~\cite{GoogleMind2019} exists in the systems ML field as well; %
it proves that for wide NN the learning dynamics simplify considerably and that in high width they are governed by a linear model. 
 Our finding gives an empirical equivalent of their mathematically proven theorem while providing a practical method to widen the NN automatically (see Sec.~\ref{HWdesign} of the paper). GoogleMind study, while proving it theoretically, has not proposed a method/tool to automatically widen the NN and make it linear. Our paper proposes an effective and automated strategy to select highly correlated internal signals of DNN and combine them into one as an input node, to design fast and HW-friendly ML for solving complicated problems.  This simplification of NN can also make DNN more interpretable. This equivalence can actually be represented mathematically but it is beyond the scope of this paper.

This paper makes the following important contributions:

\begin{itemize}[noitemsep,topsep=1pt,parsep=1pt,partopsep=1pt]
\item  Selectively triggering defenses, which reduces the overhead of state-of-the-art %
defenses such as InvisiSpec (from 27\% to 1.26\%.) and Fencing (74\% to 3.46\%), 
making them practical for deployment.  
\item {It demonstrates a robust end-to-end defense system %
for 19 categories of microarchitectural attacks.} Many have no mitigation and have not been previously detected. 
\item {It presents an automatic method for engineering new security-centric HPCs.}
\item {It provides a generative framework for %
automatically  generating %
microarchitectural samples of attacks of a given {\em type} covering a broad range of microarchitectural attacks.} %

\item {It presents a novel training method that disables powerful Adversarial ML Attack} and {defends against automatically generated attacks from fuzzing tools (23.5\% improvement)}.

\item  Shows effectiveness for {Zero-Day attack detection %
with a 10.8\% accuracy improvement over the state-of-the-art.} %

\item Proposes a new metric to verify, interpret, and visualize the state of microarcitectural attacks as well as correctness and quality of the attacks generated. %
\end{itemize}

\section{Background}
\label{sec:bkgd}

In this work we focus on detection of Microarchitectural {\em side channel} attacks %
and memory attacks such as RowHammer. %
\para{Side Channel Attacks. }
Non-Transient side-channel attacks include flush-based attacks such as Flush+Fush~\cite{GrussFlushFlush}, Flush+Reload~\cite{FlushReload2014Yarom}. Conflict-based attacks such as Prime+Probe~\cite{PrimeProbe2015last} and Evict+Time~\cite{osvik2006cache}, as well as attacks on other microarchitectural structures such as the branch predictor~\cite{aciiccmez2007yet,evtyushkin-16}. %
Transient attacks consist of a leakage gadget and a transfer gadget. The leakage  gadget can be either fault-based such as Meltdown~\cite{meltdown}, Zombieload~\cite{schwarz2019zombieload}, LVI ~\cite{vanbulck2020lvi} and Fallout~\cite{canella2019fallout}, or Speculation based such as, SpectrePHT~\cite{kocher2019spectre}, SpectreBTB(or SpectreV2)~\cite{kocher2019spectre,canella2018systematic}, SpectreRSB~\cite{koruyeh2018spectre} and NETSpectre~\cite{netspectre}. 
Transient attacks use techniques from the  non-transient attacks as their transfer gadget. 
Transient attacks have the capacity to leak {\em arbitrary memory} from a victim process and can even dump the whole kernel or victim memory. Transient attacks such as Meltdown and Spectre leverage Out of Order Execution (OOO) and Speculative Execution to leak data during the transient speculation window, %
breaking memory isolation.%

\para{Example of exploitable Performance Optimization
properties:} The Branch Predictor predicts which direction and target of a branch should be taken. The Front-end fetches the instrucions and decodes them to micro-operatons (uOP). Instructions are speculativly executed on the predicted path, enabling run ahead to fill up the idle execution units and hide memory access latency. If the branch prediction was incorrect, the ROB allows rollback to the correct state. The Scheduler allows the CPU to use data values as they become available and renames registers to solve RAW, WAR and WAW hazards. The ROB keeps track of all uOps with respect to the instruction stream. We can load caches during speculative execution. The Address Generation Unit (AGU) load and store units are directly connected to the memory subsystem to process its requests. {\em Inaccessible data is provided before interrupt flags and during wrong paths.}

\para{Transient Attack Examples.} In a Meltdown attack the attacker: 1) flushes the memory space in the cache; 2) performs a Syscall/prefetch to have the kernel address loaded in L1 cache; 3) Loads the secret to a register R; 4) fills the ROB with dependent instructions which use a different execution unit; 5) Issues a transient load to an address, indexed by the content of Register; and finally 6) Measures the time of Reload.
Main phases in Spectre-PHT are: 1) Attacker Flushes the accessible cache lines; 2) Mistrains the branch predictor to make a wrong prediction; 3) Performs a transient access to inaccessible data; and 4) Times the access to the monitored line.  Even if the processor prevents speculative execution of instructions in the user process access the kernel memory, the Spectre attack still works.

\para{Memory Attacks} -
Rowhammer compromises data integrity rather than confidentiality.
DRAM cells leak charge over time and must be periodically refreshed to prevent data corruption. 
Attackers can exploit the shared DRAM buffer to not only construct a side channel (DRAMA~\cite{pessl2016drama}) but also amplify the data corruption and cause bit flips in targeted memory pages (Rowhammer~\cite{rowhammer}), 
compromising OS and browser security~\cite{gruss2016rowhammer,frigo2018grand,gruss2018another}. Recent mitigations are shown to be not fully effective against Rowhammer~\cite{frigo2020trrespass,cojocar2019exploiting}

\para{LVI} - In LVI\cite{vanbulck2020lvi}, first the adversary injects its data. %
The victim loads this data and stores it in a microarchitectural buffer. %
The load whose translation takes long time to settle and thus the stored adversary data gets forwarded speculatively. When the recovery happens, victim's secret is inferrable by accessing microarchitectural buffers. 
Flushing shared buffers, %
software patches, microcode mitigations, %
Spectre hardware mitigations and %
constant time programming  \cite{fact} all fail to mitigate LVI.  %
The only current mitigation fences on every load which is reported to have a 900\% overhead \cite{vanbulck2020lvi}.

\para{Why Transient Attacks Can be Detected in Microarchitectural Layer.}
We observe that each attack phase leads to misuse  or under utilization of processor resources which is a direct side effect of bringing the microarchitecure into a state of leakage and recovery, making these invariant features for leakage. 
The processor has HPC's built in for debugging, verification and optimization purposes which monitor events occurring in processor resources. Affected resources include the execution engine, branch predictor, TLBs, caches, MMU caches, buses, buffers, cache coherency, power and more. Flags of activities in the system propagate up and down the pipeline (e.g., stalls or transient instructions travel down the pipeline) and can be detected by monitoring different events affecting these resources.%
\vspace{-0.1cm}\section{Related Works}
\para{Current Mitigations \& their Overhead. }{InvisiSpec} ~\cite{yan2018invisispec} adds a separate speculative buffer for placing the speculative load data until it is safe to be visible. InvisiSpec suffers from high overhead (27\%). Fencing speculative loads to defend against Spectre attacks, have over 70\% overhead. Utilizing microcode-level optimizations to surgically inject specialized fences~\cite{taram2019context} mitigates some variants of Spectre attack but does not mitigate the dangerous MDS and LVI attacks. Fencing all loads, suffers from high performance overhead of up to (900\%)~\cite{vanbulck2020lvi}. 
In this work we reduce these overheads drastically, by 95\% or more, while providing full defenses for even larger spectrum of attacks than individual patches. With reduced overhead, these mitigations become feasible, providing security in an elastic way to the microarchitecture.  

\para{SW Detectors. } Malware detectors from SW layer are unable to detect  microarchitectural attacks because microarchitectural attack signature are fundamentally different than Malware~\cite{PerSpectron}. Prior works used features related to committed state
such as instruction mixes or memory
access distribution. Detection of transient attacks requires access  transient features. 
SW can only monitor 4 HPCs at a time without multiplexing counters. This provides opportunity for attacker to evade detection. SW detectors have low sampling frequency (100ms vs ns in HW) making them vulnerable to Bandwidth evasion. \cite{Gaudiot2020} Sampling and computation happen on the critical path leading to high overhead. And thus, prior work in SW detection fails to detect stealthy transient attacks. \cite{Gaudiot2020, PerSpectron}. Therefore, in this work we focus on HW-based detection of transient attacks. 

\para{Static/Non-ML detections. }  Prior work in HW based detection has followed two general approaches: (1) Static-signature-based detection similar to Cyclone~\cite{cyclone2019} and (2) anomaly detection using ML similar to~\cite{PerSpectron}. 
Static-signature-based detection, detects static properties for example Cyclone~\cite{cyclone2019} detects contention-based cache attacks by tracking the cyclic interference property in hardware. Cyclone can only detect two types of microarchitectural cache attacks (Flush+Reload and Prime+Probe).

Cyclone detects after leakage while ML-based approaches detect attacks prior to leakage. \cite{PerSpectron} %
ML-based approach can generalize to new attacks because it is focused on {\em system properties} looking for activities that are anomalous with respect to normal system behaviour unlike static-signature-based detection. %

\para{PerSpectron. }Recent work has demonstrated success at detecting microarchitectural attacks in hardware with low performance overhead~\cite{PerSpectron}.  While software detectors are vulnerable to bandwidth evasion techniques, detection in HW is resilient to such evasions~\cite{PerSpectron}.  Due to their efficacy, manufacturers are beginning to support Hardware Malware Detectors (HMDs) to mitigate threats such as Ransomware~\cite{IntelHShield} and malicious mobile apps~\cite{QCOMMHmd}. PerSpectron utilizes a single layer perceptron in hardware to monitor 106 features spanned over the processor i.e., replicated detectors in the processor to detect seen Spectre-PHT, Spectre-RSB, Meltdown, Prime+Probe, Flush+Reload, Flush+Flush attacks and CacheOut, Spectre-BTB out of sample. 

{Perceptron is fast and HW friendly.  But can only learn linear separable functions.} This problem exacerbate when trying to learn many novel categories/classes of attacks e.g., new LVI type attacks, with one structure. This work extends the number of novel %
attacks from 7 to 19 while increasing the overall detection accuracy.  PerSpectron is also  susceptible to AML attacks. In this work we propose an adversarial hardening methods that makes PerSpectron resilient to adversarial attacks. 
ML based detectors~\cite{PerSpectron} can also be tuned to increase sensitivity to detect several microarchitectural attacks. One major issue with PerSpectron is high false positives and vulnerability to new Evasive technologies. 

\para{ML Related Studies. } Augmenting the training dataset has
been used to increase the accuracy of a model. ~\cite{Ba2014DoDN} Shows that SNN can achieve same accuracy as DNN when trained with data generated by DNN using model compression. ~\cite{GraspGan} uses GAN data augmentation for robotic grasping. ~\cite{esteban2017realvalued}  Conditional RGAN used to generate synthetic medical records for training physicians. However ~\cite{Goodfellow2015ADVexample} suggests that linear models lack the capacity to resist adversarial perturbation. We show that not to be the case. ~\cite{miyato2016distributional} shows that virtual adversarial training efficiently moves decision boundaries in the adversarial direction. ~\cite{moosavidezfooli2016deepfool} demonstrated that iterative linearization of the classifier can generate minimal perturbations that are sufficient to change classification labels. ~\cite{oliver2019realisticEvaluation} suggest SSL when there are no high-quality labeled datasets from similar domains to use for fine-tuning. ~\cite{cleverhans2018} discussed a number of methods for injecting adversarial examples during training to improve the generalization of a machine learning model. ~\cite{szegedy2014going} Google improved image recognition with deep CNN. ~\cite{cleverhans2018} used GAN to improve student/teacher network learning.  Apple, Google and other researchers have used virtual adversarial training to improve ML models~\cite{ oliver2019realisticEvaluation, killoran2017generating}.

\section{Threat model} 
We assume the following: 
(1) A machine with a Hardware Malware Detector (HMD) 
trained to detect microarchitectural attacks similar to PerSpectron~\cite{PerSpectron}. 
(2) An attacker with access to a similar detector, capable of using it to craft and test evasive attacks.
(3) We assume no specific evasion strategy; {\em i.e.}, we assume no limits on how the attacker chooses to evade provided the program is legal and the attack is preserved.  
The attacker could be using arbitrary methods to evade signature-based detectors, including those developed in the malware community~\cite{PaulKocher,kocher2019spectre}
or fuzzing based approaches~\cite{weber2021osiris,moghimi2020medusa,TRRespass}. 

\section{EVAX}
\begin{figure}[!t]
\centering
\includegraphics[width=0.5\textwidth]{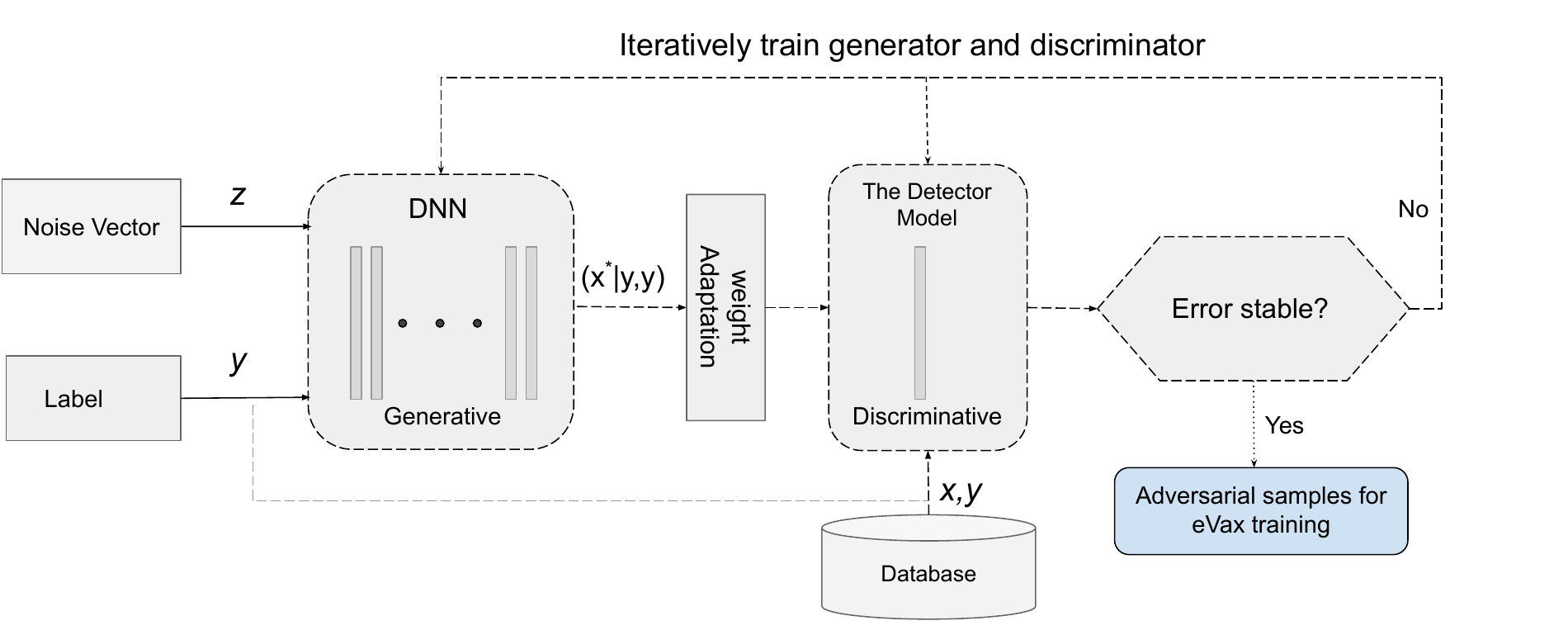}
\caption{\scheme{} AM-GAN Training Diagram} 
\label{fig:evaxsystemoverview}
\end{figure}

\begin{figure*}[!htbp]
\centering
   \begin{tabular}[t]{@{} p{\linewidth} @{}} \hline{\footnotesize
\begin{algorithmic}[1]
\Procedure{VaccinateHardwareDetector($S^\prime$)}{}   
\Procedure {AM-GAN Training}{ }
\While { Generator is improving}
\For  {\text{each training iteration}}
\State {$\text{[\textit{x}, \textit{c} , \textit{t}]} \gets \text{get\_sample [\textit{S}]} $} \Comment{\textit{x} is a sample} \Comment{\textit{t} $\in$ target malicious$/$safe}
\Comment{\textit{c} is attack's type} \Comment{$S$ is the training set}
\Procedure{Discriminator Training}{ }\Comment{Train the discriminator to distinguish between real and generated data.}

\State {$ $y$ \gets \text{Discriminator ($x, c$)}$}\Comment{\textit{y} is $1$ for \textit{real and matching} , $0$ for \textit{fake}}
\State $\textit{$d_x$} \gets \text{\textit{t-y}}$\Comment{$d_x$ is discriminator's error} \label{op:error}

\State $ ${z}$ \gets \text{RandomNoise(145)}$\Comment{\textit{z} is a noise vector for 145 features}
\State {$\textit{$x^{\star}$} \gets \text{Generator (\textit{z},\textit{c})}$}\Comment{\textit{$x^{\star}$} is adversarial$/$fake example}
\State {$\textit{$y^{\star}$} \gets \text{Discriminator ($x^{\star}$)}$}\Comment{$y^{\star}$ is $1/0$ for \textit{real and matching}$/$\textit{fake}}
\State $\textit{$d_{y^{\star}}$} \gets \text{\textit{$y^{\star}$-y}}$\Comment{$d_{y^{\star}}$ is discriminator's error} \label{op:errorD}
\State $\text{Update(Discriminator , $d_{y^{\star}}+d_x$ , c) }$\Comment{Update the discriminator's weights seeking to minimize  $d_x+d_{y^{\star}}$ }
\EndProcedure

\Procedure {Generator Training}{ }\Comment{Train the generator to output data that "fools" the discriminator}

\State $\textit{z} \gets \text{RandomNoise(145)}$\Comment{\textit{z} is a noise vector for 145 features}
\State {$\textit{$x^{\star}$} \gets \text{Generator (\textit{z} , \textit{t} , \textit{c})}$}\Comment{\textit{$x^{\star}$} is adversarial$/$fake example}

\State {$\textit{y} \gets \text{Discriminator ($x^{\star}$)}$}\Comment{\textit{y} is $0/1$ for \textit{fake, unmatching}$/$\textit{real and matching pairs}}
\State ${d_{x^{\star}}} \gets \text{\textit{t-y}}$
\Comment{ $d_{x^{\star}}$   is Discriminator's error} 

\State $\text{Update(Generator , $d_{x^{\star}}$) }$\Comment{Update the generator's weights seeking to maximize   $d_{x^{\star}}$}

\EndProcedure
\EndFor 
\EndWhile
\EndProcedure
\Procedure{Automatic Attack Generation}{$c^\prime,t^\prime)$}
\State $l \gets [c^\prime,t^\prime]$ \Comment{$t^\prime$ $\in$ desire target malicious$/$safe}
\Comment{$c^\prime$ is desired attack's type, $0$ if safe}
\State $\textit{z} \gets \text{RandomNoise(145)}$\Comment{\textit{z} is a noise vector for 145 features}
\State  $x^\prime \gets Generator (l,z) $ \Comment{Generate an attack for label $l$, $x^\prime$ is the generated attack sample} 
\State  $S^\prime \gets add(x^\prime,c^\prime,t^\prime) $  \Comment{$S^\prime$ is the augmented training set}

\For {\text{each training iteration}}\Comment{Generate a batch of new attacks}
\State $[c^\prime,t^\prime]  \gets \text{get\_label [\textit{S}]}$\Comment{Iterate through label list}
\State $Automatic Attack Generation(c^\prime,t^\prime)$\Comment{Generate new attack for type $c^\prime$ and target $t^\prime$ }

\EndFor
\EndProcedure
\EndProcedure

\end{algorithmic}
}

\end{tabular}\hfill
\vspace*{-2mm} 
\caption{\scheme{} Training algorithm.}
\label{fig:algo}
\end{figure*}

\subsection{Training Generative Modeling Framework}

\para{The Goal. }The goal of \scheme{} is to develop a detector that generalizes across a large number of seen and unseen microarchitectural attacks and is also resilient to evasion.

Recall that adversarial algorithms aim to generate a single example that fools the detector for a given input and potential target class; typically, the algorithm determines an effective perturbation pattern that guides the input across the classification boundary to cause it to be misclassified.  

We would like to automatically generate attack samples that represent the feature map for these undisclosed or undiscovered variations of attacks. These samples will harden our detector and %
sharpen its classification boundaries making it resilient to evasion.

\medskip
\para{Our Solution. }To solve this problem, we use a training process with a 
(GAN)-based game. 
 The idea is that since GANs can generate realistic instances of data that no one has seen, vector representations of these instances can also be used to describe evasive or undisclosed variations of attacks.
 GANs improve the training process by turning the attack generation problem into a {\em self supervised} learning problem between two sub-models, a Generator and a Discriminator.
The two sub-models are trained together in an {\em adversarial game}. The Generator learns the domain specifics automatically through playing the game with the Discriminator instead of training on limited data. 
In a traditional attack detector, the classifier learns from input/output data; therefore, the classification ability is limited to seen data points. But GAN does not learn from an
input/output data mapping to tell the model how it should generate an attack.
The Discriminator, in contrast to the classifier in conventional detectors, learns how inputs and outputs can be paired in a  microarchitectural system, and it tells the model whether it was correct. 
Because of this, the Generator ultimately learns a broad range of the possible correct answers

This is an important principle of our design and the key reason for ~\scheme{}'s high effectiveness.

\para{The Game Setup.} 
Our Discriminator in this game has the architecture of our hardware detector. We start with a single layer perceptron-based classifier  similar in architecture to PerSpectron~\cite{PerSpectron}; however, this
technique would generalize to any ML-based detector.
Our Generator is a deep neural network, which creates 
an unusual asymmetry between the Generator and Discriminator.  We will label this
an Asymmetric Model GAN (AM-GAN).  
As the adversarial game unfolds, the Generator learns each time a sample fools the Discriminator. 
Instead of recognizing a pattern, the Generator learns to create attack samples essentially from scratch; indeed, the input into the Generator is mainly a (noise) vector of random numbers, the feedback from the Discriminator and the label. 
Conversely, each time the Discriminator rejects an attack sample, the Generator uses the feedback to improve its strategy. 
Even though the Generator always generates samples without the knowledge of seen attacks (ground truth), the overall training of our AM-GAN is supervised because real attack samples are used to train the Discriminator. 
As the Generator gets better at producing realistic attack samples, the Discriminator gets better at telling generated samples from original data.
Moreover, as the Discriminator gets better at detecting a certain evasion strategy, the Generator is forced to explore additional strategies, exposing the Discriminator to them, further improving its robustness. Both models continue to improve simultaneously.

\para{Integrating Labels.  }Our AM-GAN (Figure~\ref{fig:overview}) can also determine the kind of attack sample it generates. 
In a simple GAN, there is no means to control the attack type; utilizing a conditional GAN (CGAN) allows us to customize properties of samples for each attack class.
Our Discriminator learns to accept matching pairs from the seen database while rejecting pairs that are mismatched and pairs that are produced by the Generator. 
Our final Generator is capable of producing realistic microarchitectural attack samples for each attack type we give a label to. We enter the descriptive features of an attack into our Generator and have it output a range of matching samples, expediting the process of adversarial attack sample generation. In Section~\ref{sec:verify} we introduce a metric to measure the semantic loss for generated attacks and visually verify this capability of our AM-GAN.

 \subsection{Attack Sample Generation} 
 
\begin{figure}[!t]
\centering
\includegraphics[width=0.5\textwidth]{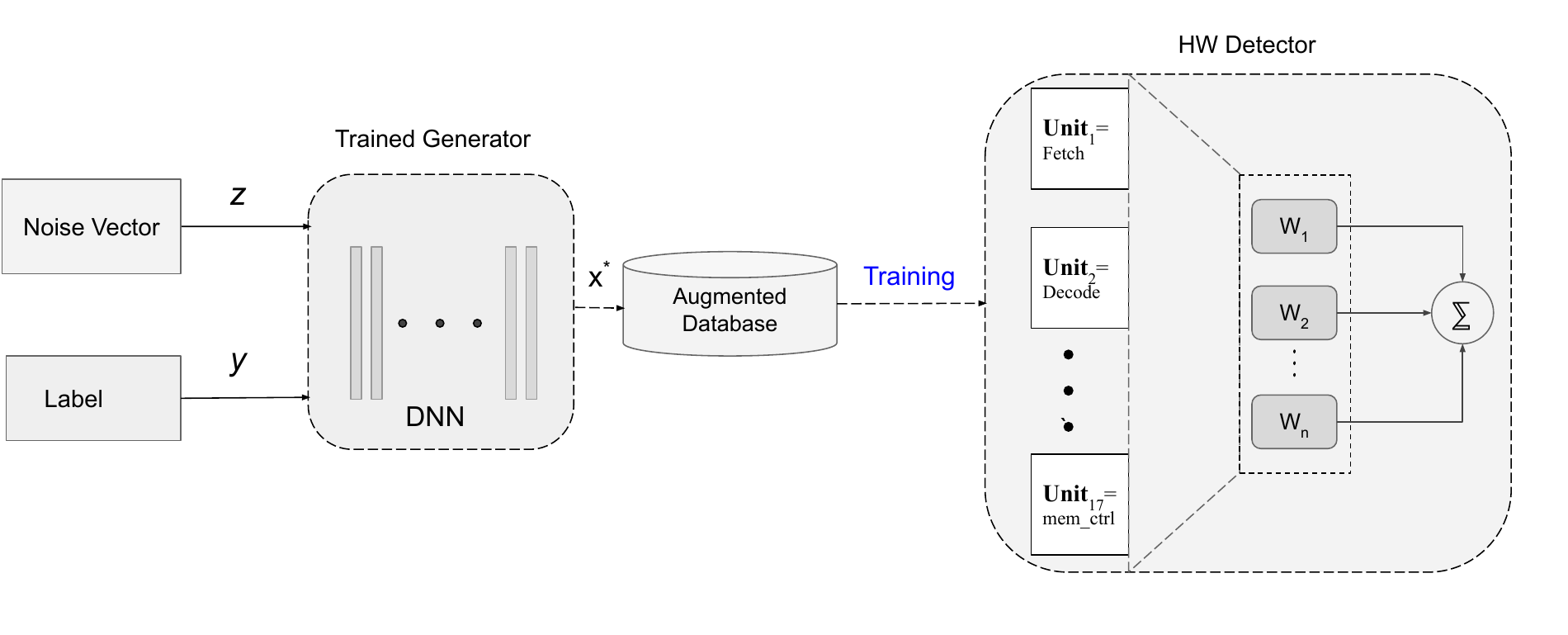}
\vspace{-0.85cm}\caption{\scheme{} HW Detector Training Diagram} 
\label{fig:overview}
\end{figure}

 Figure~\ref{fig:overview} depicts a block diagram of our AM-GAN architecture and it's training process (see Algorithm in Figure~\ref{fig:algo}).
In this section we explain the detailed steps of our training algorithm. 
 \vspace*{0.5mm}

\para{Training the Generator:}
For each training iteration:

\begin{CompactItemize}
\item  The Generator takes a new random noise vector z and an attack type label
{\em e.g., spectre-RSB-type attack}. 
The Generator generates an example $x^{\star}$ that strives to be both an adversarial attack ({\em e.g.}, looks like a safe program) and a convincing match for its label. %

\item The Discriminator network classifies $x^{\star}$ into 0/1. 
One indicates a match with seen sample-labels from the dataset and zero indicates unmatched sample-labels and generated sample-labels.

\item The classification error of the Discriminator will be computed and backpropogated to update the Generator weights, seeking to maximize the Discriminator's error.

\end{CompactItemize}

\para{Training the Discriminator:}
For each iteration:
\begin{CompactItemize}
\item We choose a seen program sample $x$ from the training dataset with its associated label.

\item Generator gets a new random noise vector $z$ and label describing the attack type and synthesizes an adversarial example $x^{\star}$.

\item Discriminator receives the seen example $x$ and label, %
the adversarially generated sample-label $x^{\star}$ and the label that was used to generate them.
\item For both examples the Discriminator outputs a probability indicating whether the input example was seen data from the dataset, matching its label pair 
or unmatched pairs.

\item The classification error of Discriminator is computed and backpropogated to update its weights, seeking to minimize classification error. 
\item The training ends with our AM-GAN reaching Nash equilibrium, meaning that at this point the Generator produces fake examples indistinguishable from seen attacks in the training data, and the Discriminator can at best randomly guess whether a particular example is a seen attack or a generated one. 
\end{CompactItemize}

\begin{figure}[!htbp]
\centering
\includegraphics[width=0.5\textwidth]{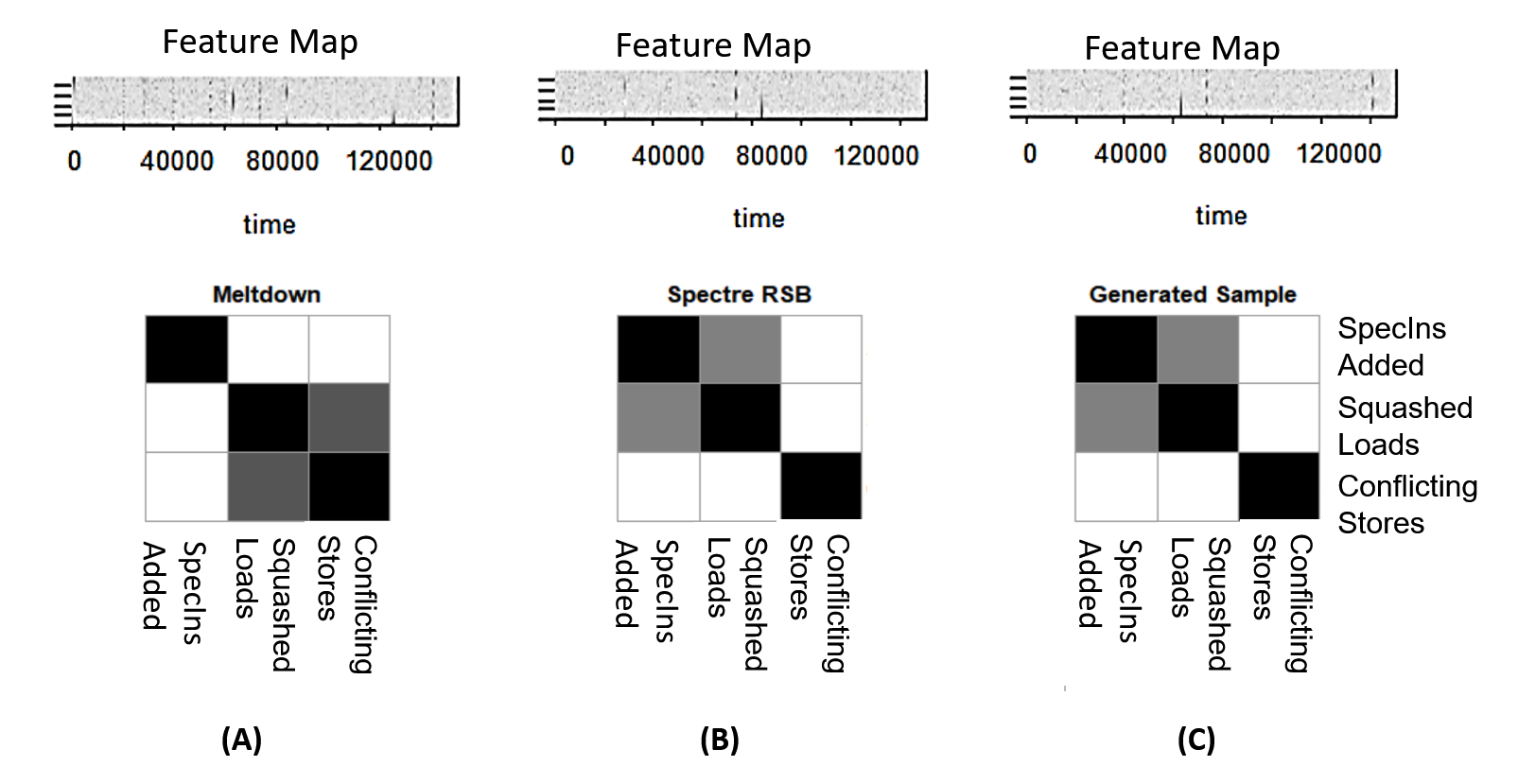}
\vspace*{-0.5cm}
\caption{Part of the Gram matrix  during leakage phase for two attacks and an attack sample generated by our framework with the label SPECTRE-RSB -- the darker color represents larger values. 
Attacks (B) and (C), similar in type, have similar Gram matrices. Even though the values of the features may be very different, the Gram matrix--a measure of the correlation between pairs of features--is similar.  }

\label{fig:Gram}
\end{figure}

\subsection{Training the Hardware Detector }

It is important not to confuse our  Discriminator with our final Detector. 
We emphasize that we can not use this Discriminator as the attack detector directly. While they are both classifiers, 
the Discriminator adapts to answer a different, although related, question: whether a generated attack sample is {\em fake} or {\em real} (seen attack or generated).  It does not perform well as a Detector whose goal is to classify {\em suspicious} vs. {\em safe} activity. 
 After reaching a stable error (it is hard to achieve Nash equilibrium) we perform the process of hardening our actual detector using AM-GAN: 
\begin{CompactItemize}
\item We feed the trained Generator with each attack type from the training set (conditioning inputs) and collect the generated examples, using them to augment the training database. 
The examples produced by our Generator are adversarial attack samples that carry the footprints of microarchitectural attacks but are very hard to distinguish from benign programs. 
\item We retrain our hardware detector with the augmented database. The generated examples which consistently fool the Discriminator are used to train our ~\scheme{}, thereby hardening it against such evasive attacks.

\end{CompactItemize}

\subsection{Interpretability \& Quality measure for the Generated Samples
}\label{sec:verify}

 In AM-GAN, just by looking at the two loss functions it is unclear when we've actually finished training, since when one gets better the other one gets a larger loss. 
 In principle, Nash equilibrium gives us a reasonable stopping criteria but in practice reaching equilibrium is difficult.
To cover the adversarial space, we would want to start collecting training data from the generator when the generated examples are different enough from the seen data in values but belong to the same class.  As explained in Section~\ref{sec:bkgd},
several basic blocks represented as misused or under-utilization are required to be present simultaneously for attacks to build a side channel. So we want to score samples based on the presence and approximate values of low-level microarchitectural states required for successful construction of a channel, leakage, and recovery of each type.

Our solution is based on the observation that while one attack and its evasive version have different distribution of instructions they share commonalities in micro-arhitectural state during different phases of the attack.
We want every new attack generated by AM-GANs Generator, conditioned by an attack type (label), to have some common correlation patterns during leakage between features (mutual correlation pattern) with the seen example of the attack category. So we define a \textit{Gram matrix} that contains the dot product between all possible feature pairs.

\para{Gram Matrix. } 
To numerically measure how often two feature maps are present together,  we multiply the values of two vectors in each position and sum the results. If the resulting value is high, the features are highly correlated.  
Figure~\ref{fig:Gram} shows the Gram matrices for three features for each of the attacks. 
The ribbons on the top of Figure~\ref{fig:Gram} are the program map.
Each pixel shows the value of the microarchitectural features over time. The bottom part of the Figure shows {\em correlations} between three chosen features.
\para{Microarchitectural Leakage Snapshot (Visualization). } For example, to generate the attack (C) in Figure~\ref{fig:Gram} we have fed the model with a \textit{spectre-RSB} condition. We were able to visually check that \scheme{} is working as expected since the Gram matrices for the three features in attack (B) and (C) match but (A) and (C) mismatch since attack (A) is meltdown and (B) is spectre. 
While {\em leakage style} is similar, the feature maps on the other hand for attack (B) and (C) are different
-- verifying that AM-GAN found another variation of the same attack type matching a different binary. 

\para{Interpretability of Features. } Visualization of AM-GAN generated samples also helps the designer to identify and interpret the highly correlated features. For example, we can see that the features \textit{Conflicts in Instruction Queue} and \textit{SquashedLoads}  fire strongly together in Meltdown but not in Spectre-RSB and the generated attack (C). 
Conversely \textit{Speculative Instructions Added} often fire strongly together in Spectre-RSB and the generated attack (C).  
We can visually see that Meltdown-type  samples rely on Conflicts in Instruction Queue (OoO execution). We see that squashed loads in Speculative attack samples generated are caused by speculation not exceptions. These examples verify that our AM-GAN framework is trained long enough to represent the continuous target distributions of attacks from which we can sample for training EVAX. 
This is a powerful tool that can be used for further analysis and tuning of the features of the final product and for developing mitigations of future attacks.  

\begin{figure}[!htbp] 
\centering
\includegraphics[width=0.27\textwidth]{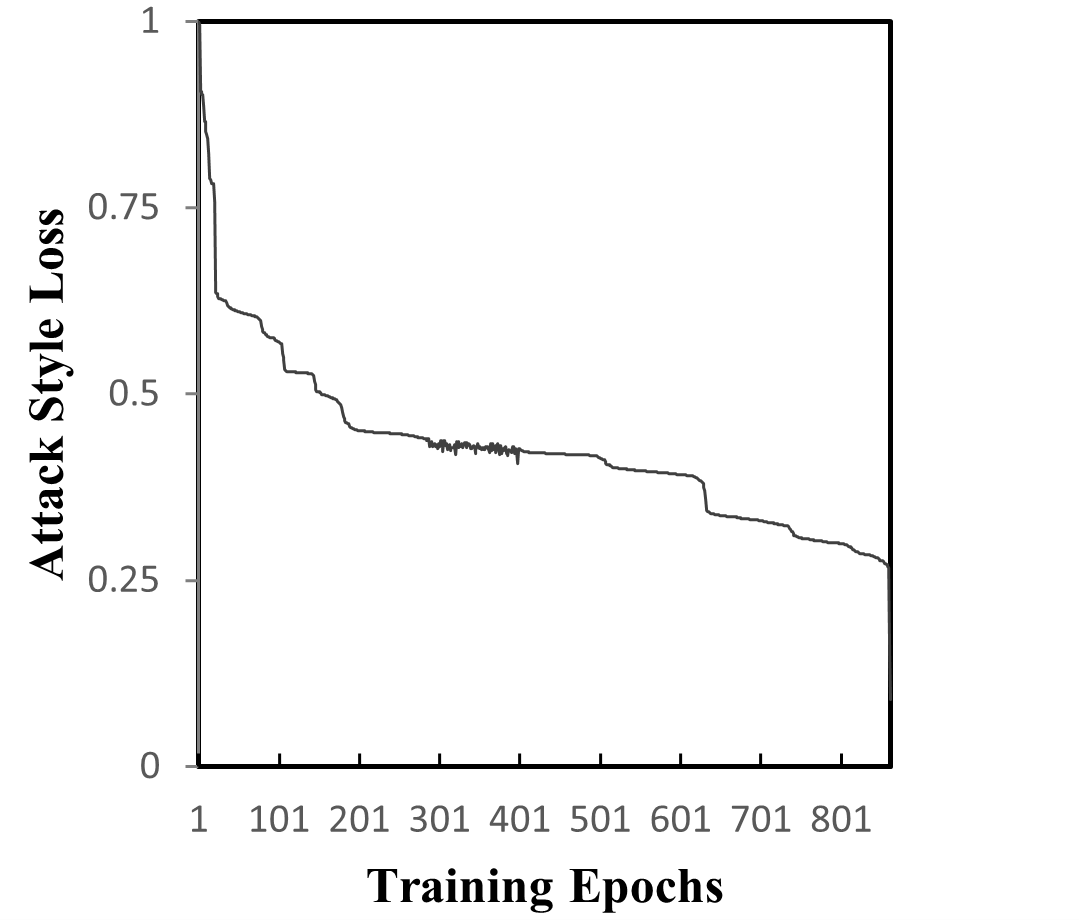}
\vspace*{-1mm}
\caption{
 Attacks Style Loss during AM-GAN training.
}
\label{fig:style}
\end{figure}
\para{Attack Quality Measure. }To calculate the attack style loss, we calculate the Gram matrix (GM) for a set of features for the base and generated attacks and then compare their similarity using sum of squared errors. Attacks (B) and (C) which are similar in type, will have similar Gram matrices resulting in leakage style loss near zero. 
Algebraically, the attack leakage style loss $(L_{GM})$ between the base attack (B) and generated attack (G), can be written as follows:

{\footnotesize
\[
L_{GM}(B,G)=\frac{1}{4{\alpha}N^{2}}\sum_{ij} (GM(B)_{ij} - GM(G)_{ij})^{2}
\]
}
Where $N$ is the number of features and $\alpha$ is a constant. 

In Figure~\ref{fig:style}
we can see that attack footprint style loss $(L_{GM})$ reduces 
with each training iteration--the output becomes stylistically 
closer to the type of the attacks that it is labeled to produce --- meaning as the training epoch increases, the {\em quality} of the generated attacks gradually becomes higher. 

We monitor this metric as AM-GAN is being trained and start collecting training data when 
the absolute value of the $(L_{GM})$ for our  generated examples becomes quite small $( 0.1\pm0.006)$ indicating the sample represents the microarchitectural state of leakage related to the attack type. We visually verified the quality and semantic consistency of one example generated of each attack type in our database before starting to collect samples to augment our dataset.

\section{HW Design}\label{HWdesign}

\subsection{Feature Engineering}
{We want to achieve  accuracy comparable to a deep neural network with a Security Friendly (fast classification) \& HW Friendly (light) model (shallow/linear neural network)}.
Simple models can provide classification fast enough to detect transient attacks before leakage.  However, we also
find that such simple models are capable of generating highly accurate predictions.

\begin{figure}[!htp]
    \centering
    \includegraphics[width=0.35\textwidth]{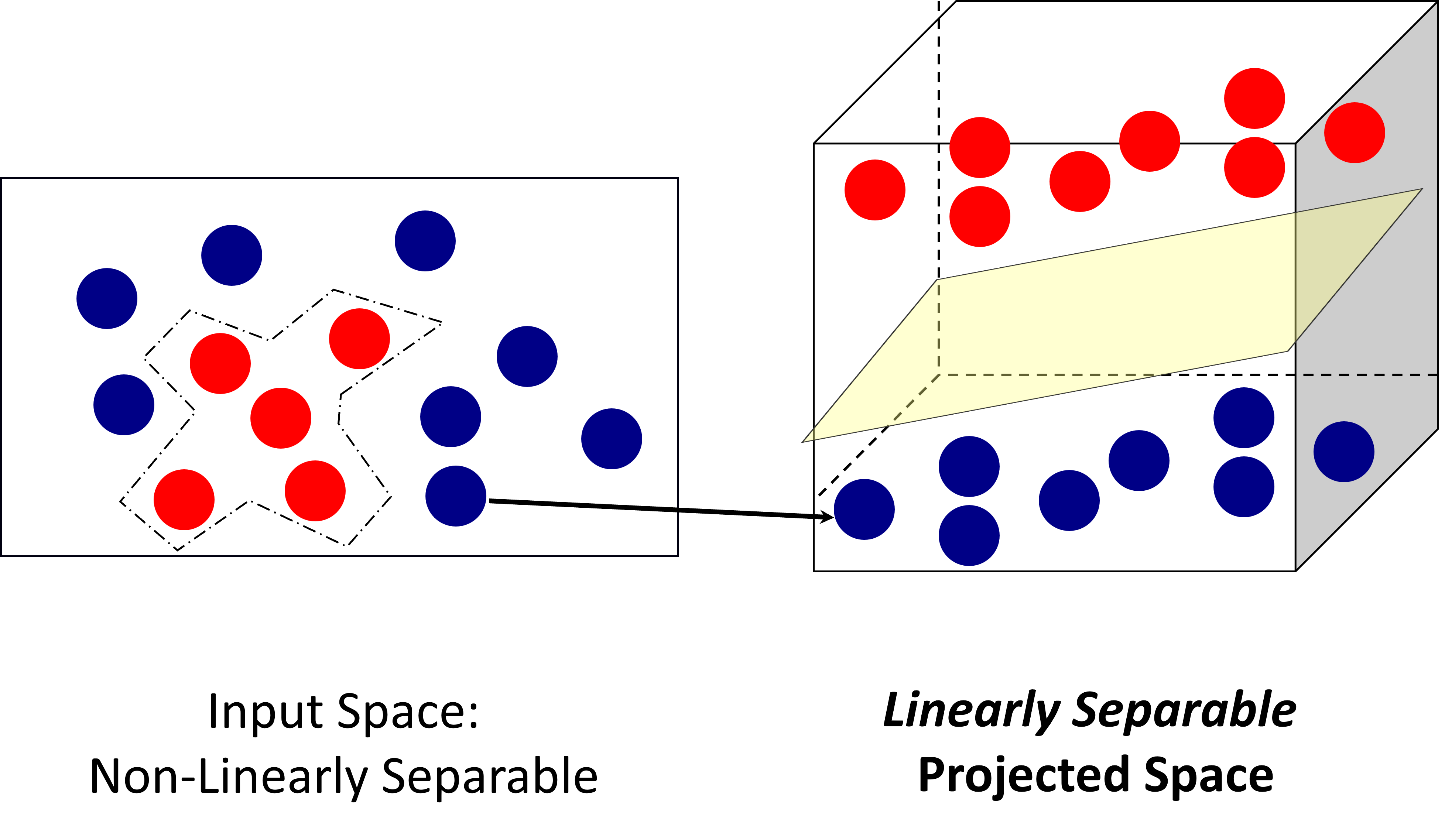}
    \caption{Effect of added dimension (new performance counters for security) on classification.  }
    \label{fig:higherD}
\end{figure}

\para{Solving a multi dimensional problem with a simple architecture. }  We want to use the same simple (e.g.,  perceptron) classifier PerSpectron uses
to classify many more categories of novel attacks that the previous model fails to learn, such as LVI-type. 
Our solution is to add Dimension to the input space. Introducing new performance counters and monitoring a large set of  microarchitectural features
transforms a Non-Linearly Separable Space to Linear. 
This allows a simpler architecture — no hidden layer is necessary.  
This is because the capacity of an ML model depends on both input space and the model itself. 
As illustrated in figure~\ref{fig:higherD} higher dimension enables a simpler classifier. It seems that there is a constant total value of  {\em classification Volume} which can be captured by the dimension of input space multiplied by order of classification function (complexity); if you increase the input dimensions, you can use a simpler classifier.
In order to expand the detection ability of the prior works to  more categories of novel attacks, we include more HW performance counters (145 instead of 106) including 12 newly engineered security-centric performance counters. These new engineered features enabled the same linear model trained for classifying 6 categories of attack to also learn 19 other classes of attacks.  
Additionally, the new security performance counters are created by an automated method of selection and combining multiple signals originated from the performance counters into one new, highly correlated counter for {\em security}. Conducting this search by brute-force requires simulation time that is intractable.
For example to simply choose three counters to combine out of 1160 one must test 259,476,920 possible combinations as most combination does not correlate to the security anomalies.  We automated this selection using AM-GAN which can select any number of useful counters (1,2,3,4,...) without the need of running all combinations.  
We will explain this method (collapsing independent counters into a smaller set and linearizing the network while maintaining the accuracy) next. 

\begin{figure}[!htbp]
    \centering
    \centering\scalebox{0.65}{%
\centering\begin{tikzpicture}[node distance=2cm]

  \begin{scope}[local bounding box=diagram]

\node [draw,	minimum width=2cm,	minimum height=1.2cm]  at (0,0) (pp) {\footnotesize Prime+Probe};

\node [draw,	minimum width=2cm,	minimum height=1.2cm, below=0.5cm of pp]  (FR) {\footnotesize Flush+Reload};

\node [draw,	minimum width=2cm,	minimum height=1.2cm, below=0.5cm of FR]  (FF) {\footnotesize Flush+Flush};

\node [draw,	minimum width=2cm,	minimum height=1.2cm, right=1.5cm of pp
]  (CH) {\footnotesize Cache Hit};

 \draw[-stealth] (pp.east) -- (CH.west) node[midway,above]{};

\node [draw,	minimum width=2cm,	minimum height=1.2cm, right=1.5cm of FR
]  (CM) {\footnotesize Cache Miss};
 \draw[-stealth] (FR.east) -- (CM.west) node[midway,above]{};
 
\node [draw,	minimum width=2cm,	minimum height=1.2cm, right=1.5cm of FF
]  (ST) {\footnotesize \textcolor{red}{Stealthy}};
 \draw[-stealth] (FF.east) -- (ST.west) node[midway,above]{};
\node [draw,
	minimum width=2cm, 
	minimum height=1.2cm,
	right=1.0cm of CM
] (NR) { $\frac{Not Reused}{Eviction}$ };
 \draw[-stealth] (CH.east) -- (NR.west) node[midway,above]{};
 \draw[-stealth] (CM.east) -- (NR.west) node[midway,above]{};
  \draw[-stealth] (ST.east) -- (NR.west) node[midway,above]{};

\node [draw,
	minimum width=2cm, 
	minimum height=1.2cm,
	right=1cm of NR
] (CE) { \textcolor{ao}{Clean Evicts}};
  \draw[-stealth] (NR.east) -- (CE.west) node[midway,above]{};

 \end{scope}  
 
  \draw[->] (CE) |- ([yshift=-1em]diagram.south) -| (ST) node[near end, below]{\textcolor{blue}{Solved}};

\end{tikzpicture}

  }
    \vspace{-1cm}    \caption{Complex HPCs for detection of stealthy cache attacks. }
    \label{fig:cacheattacks}
\end{figure}
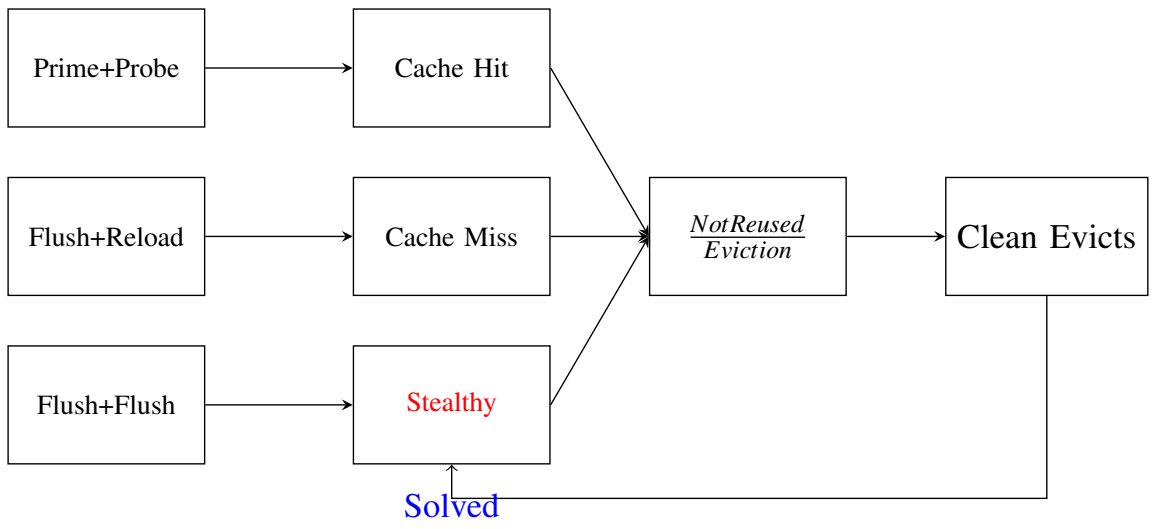

\para{Automating Performance Counter Engineering with GANs for Security.} Performance Counters are designed for tuning performance. Some of these performance counters are correlated with security. Prior work~\cite{PerSpectron} identified  (manually) complex performance counters such as Number of Clean Evicts, that was shown to be useful in detection of stealthy cache attacks (see Figure~\ref{fig:cacheattacks}). 

\begin{figure}[!htbp]
    \centering
    \centering\scalebox{0.65}{%
\centering\begin{tikzpicture}[node distance=2cm]

  \begin{scope}[local bounding box=diagram]

\node [draw,	minimum width=2cm,	minimum height=1.2cm]  at (0,0) (pp) {\footnotesize Spectre PHT};

\node [draw,	minimum width=2cm,	minimum height=1.2cm, below=0.5cm of pp]  (FR) {\footnotesize Spectre BTB};

\node [draw,	minimum width=2cm,	minimum height=1.2cm, below=0.5cm of FR]  (FF) {\footnotesize Spectre RSB};

\node [draw,	minimum width=2cm,	minimum height=1.2cm, below=0.5cm of FF]  (Melt) {\footnotesize Meltdown};

\node [draw,minimum width=3cm,	minimum height=1.2cm, right=0.5cm of pp
]  (CH) {\footnotesize {\begin{tabular}{c} DirectBrMiss\\ \textcolor{red}{Stealthy} \end{tabular}} };

 \draw[-stealth] (pp.east) -- (CH.west) node[midway,above]{};

\node [draw,	minimum width=3cm,	minimum height=1.2cm, right=0.5cm of FR
]  (CM) {\footnotesize
{\begin{tabular}{c} IndirectBrMiss \\ \textcolor{red}{Stealthy} \end{tabular}}};
 \draw[-stealth] (FR.east) -- (CM.west) node[midway,above]{};
 
\node [draw,	minimum width=3cm,	minimum height=1.2cm, right=0.5cm of FF
]  (ST) {\footnotesize
{\begin{tabular}{c} RASIncorrects \\ \textcolor{red}{Stealthy} \end{tabular}}};

\node [draw,	 minimum width=3cm,	minimum height=1.2cm, right=0.5cm of Melt
]  (melt2) {\footnotesize
{\begin{tabular}{c} SpeculativeLDs \\ \textcolor{red}{Stealthy} \end{tabular}}};

 \draw[-stealth] (Melt.east) -- (melt2.west) node[midway,above]{};

 \draw[-stealth] (FF.east) -- (ST.west) node[midway,above]{};
\node [draw,
	minimum width=2.4cm, 
	minimum height=1.2cm,
	right =1cm of CH, yshift=-1cm
] (NR) { $\frac{SquashedLoads}{SpecLoads}$ };
 \draw[-stealth] (CH.east) -- (NR.west) node[midway,above]{};
 \draw[-stealth] (CM.east) -- (NR.west) node[midway,above]{};
  \draw[-stealth] (ST.east) -- (NR.west) node[midway,above]{};

\node [draw,
minimum	 width=3.8cm, 
	minimum height=1.5cm,
	right=0.5cm of NR
] (CE) { \textcolor{ao}{\begin{tabular}{c}\footnotesize Committed Maps \footnotesize {Undone}\end{tabular}}};
  \draw[-stealth] (NR.east) -- (CE.west) node[midway,above]{};

\node [draw,
	minimum width=2.4cm, 
	minimum height=1.2cm,
	right=1cm of ST
] (NR2) { $\frac{SpecLoadHitsInWrQ}{SquashedLoad}$ };
 \draw[-stealth] (CH.east) -- (NR2.west) node[midway,above]{};
 \draw[-stealth] (CM.east) -- (NR2.west) node[midway,above]{};
  \draw[-stealth] (ST.east) -- (NR2.west) node[midway,above]{};
    \draw[-stealth] (melt2.east) -- (NR2.west) node[midway,above]{};

\node [draw,
minimum	 width=3.8cm, 
	minimum height=1.5cm,
	right=0.5cm of NR2
] (CE2) { \textcolor{ao}{\begin{tabular}{c}\footnotesize Squashed Bytes Read \\\footnotesize  from WR Queue \end{tabular}}};
  \draw[-stealth] (NR2.east) -- (CE2.west) node[midway,above]{};

 \end{scope}  

   \draw[-stealth] (CE.north) |- ([yshift=1em]diagram.north)  -| (CH.north) node[near end, below]{\textcolor{blue}{Solved}};  
  
   \draw[->] (CE2) |- ([yshift=-1em]diagram.south) -| (melt2) node[near end, below]{\textcolor{blue}{Solved}};

\end{tikzpicture}

}
    \vspace{-1cm}\caption{Complex HPCs for detection of stealthy Speculative and Meltdown type attacks. }
    \label{fig:automated}
\end{figure}
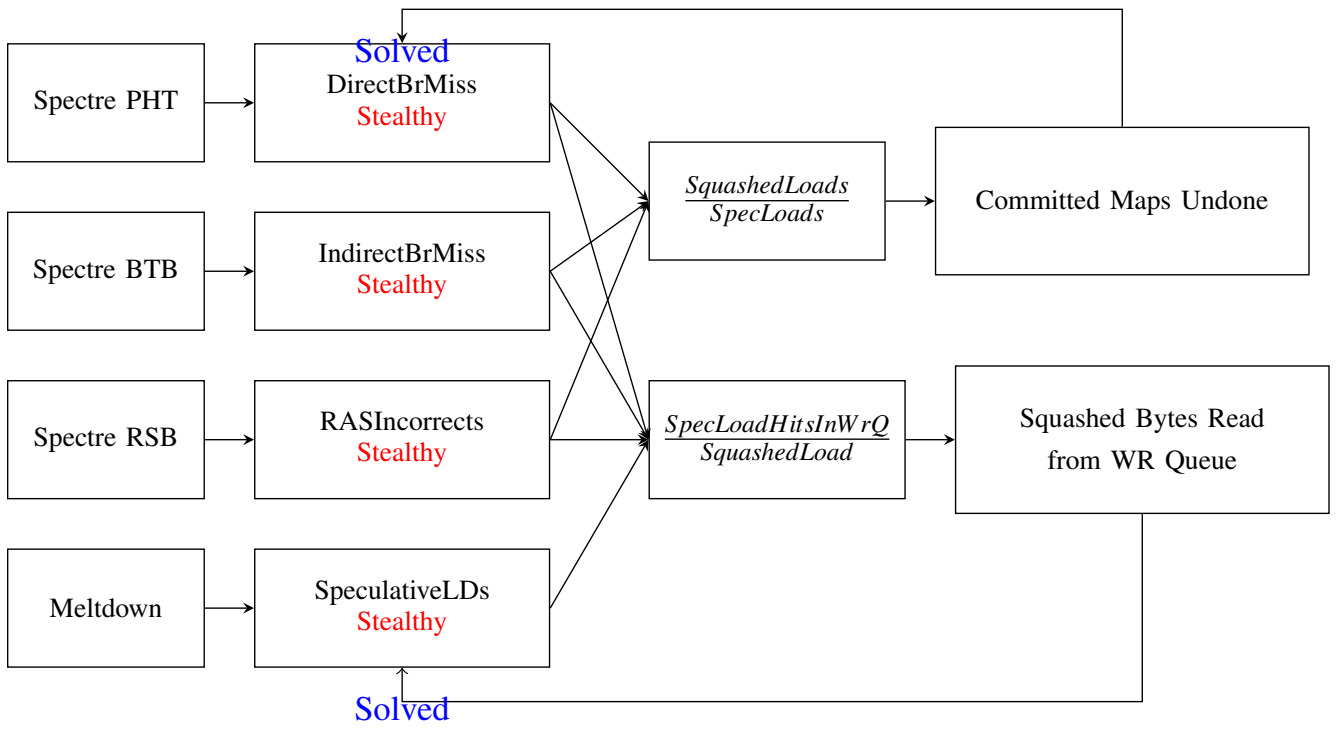

In this work 
we use the hidden nodes from our trained AM-GAN Generator to {\em automatically} engineer new counters for security. The hidden nodes receive the information from the upper layer and processes it. Then, the obtained value is sent to the next layer. The output layer will also process the information from the hidden layer and give the output. The top weights with largest absolute value dominate the instantiation of each layer. We look at hidden nodes that are connected to the first layer.  We sort the weights of the hidden layer of the network and select the top 12 nodes connected to the input HPCs. We then define the Boolean AND Logic of connected HPCs to that node as a new HPC specifically engineered for Security.  These are different than 133 highly correlated features that we already use (27 more than prior work). See figure ~\ref{fig:automated}.

\begin{figure}[!htbp]
    \centering
    \centering\scalebox{0.65}{%

\centering\begin{tikzpicture}[node distance=2cm]

  \begin{scope}[local bounding box=diagram]

\node [draw,	 	minimum width=2cm,	minimum height=1.2cm]  at (0,0) (pp) {\footnotesize CacheOut};

\node [draw,	minimum width=2cm,	minimum height=1.2cm, below=0.3cm of pp]  (FR) {\footnotesize LVI};

\node [draw,	minimum width=2cm,	minimum height=1.2cm, below=0.3cm of FR]  (FF) {\footnotesize FallOut};

\node [draw,minimum width=3cm,	minimum height=1.2cm, right=0.5cm of pp
]  (CH) {\footnotesize {\begin{tabular}{c} Squashed Load\\ \textcolor{red}{Stealthy} \end{tabular}} };

 \draw[-stealth] (pp.east) -- (CH.west) node[midway,above]{};

\node [draw,	minimum width=3cm,	minimum height=1.2cm, right=0.5cm of FR
]  (CM) {\footnotesize
{\begin{tabular}{c} lsq.forwLoads \\ \textcolor{red}{Stealthy} \end{tabular}}};
 \draw[-stealth] (FR.east) -- (CM.west) node[midway,above]{};
 
\node [draw,	minimum width=3cm,	minimum height=1.2cm, right=0.5cm of FF
]  (ST) {\footnotesize
{\begin{tabular}{c}  ConflictingLDSt\\ \textcolor{red}{Stealthy} \end{tabular}}};

 \draw[-stealth] (FF.east) -- (ST.west) node[midway,above]{};
\node [draw,
	minimum width=2cm, 
	minimum height=1.2cm,
	right=1cm of CM
] (NR) { $\frac{SpecLoadsHits}{SquashedLoad}$ };
 \draw[-stealth] (CH.east) -- (NR.west) node[midway,above]{};
 \draw[-stealth] (CM.east) -- (NR.west) node[midway,above]{};
  \draw[-stealth] (ST.east) -- (NR.west) node[midway,above]{};

\node [draw,
	minimum width=2cm, 
	minimum height=1.2cm,
	right=0.5cm of NR
] (CE) { \bf\textcolor{purple}{\begin{tabular}{c}\footnotesize Squashed ReadWrQ \end{tabular}}};
  \draw[-stealth] (NR.east) -- (CE.west) node[midway,above]{};

 \end{scope}  
 
  \draw[->] (CE) |- ([yshift=-1em]diagram.south) -| (ST) node[near end, below]{\textcolor{blue}{Solved}};

\end{tikzpicture}
}
 
    \vspace{-1cm}    \caption{Complex HPCs automatically engineered capable of detecting unseen MDS-type and LVI  attacks. }
    \label{fig:LVIHPC}
\end{figure}
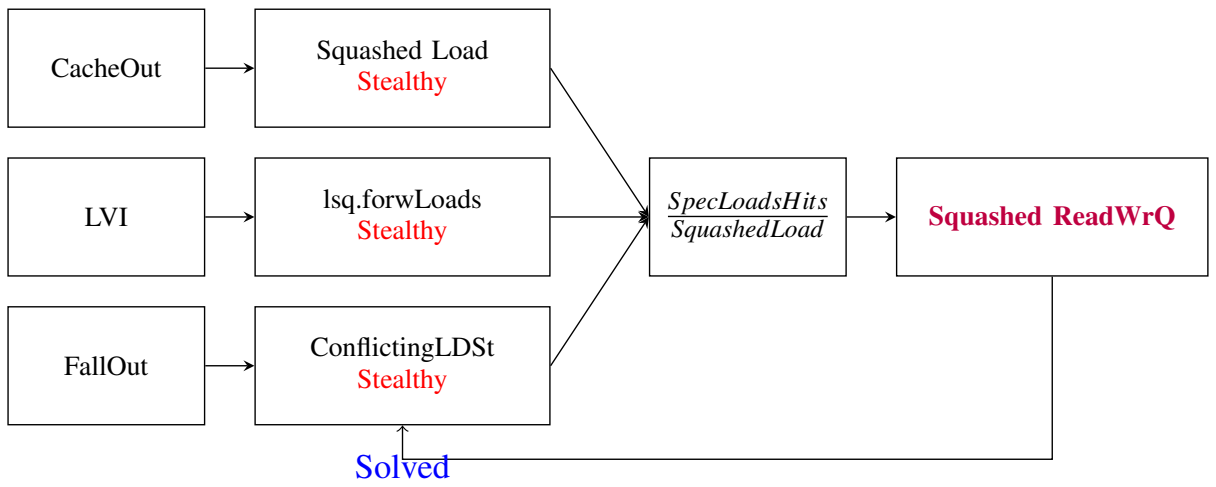

An example of these engineered features is {\em SquashedBytesReadFromWRQu}. This feature is equivalent to a node in the third layer of our AM-GAN generator combining the signals from two nodes (1) {\em Spec Loads Hit in Write Queue} and (2) {\em number of Squashed Loads} with heavy weights. {\em Spec Loads Hit in write Queue} (itself a combination of two HPCs) and {\em Number of Squashed Loads} with heavy weights are already availble HPCs. We then merge these two signals into one to engineer a new signal we call {\em SquashedBytesReadFromWRQu}, which can easily be implemented in the CPU with minimal logic. We found this new HPC specifically exposes security vulnerabilities in the transient domain. For instance, as shown in Figure~\ref{fig:LVIHPC}, we show that this new HPC detects both MDS and LVI attacks. %

\para{Replicated Feature Detector. }{ We also use the entire space by replicating mutually correlated features. } Prior work suggested use of {\em Replicated detectors} to simplify the problem of detection for perceptron-based detectors. The idea is to not disregard the mutually correlated (redundant) features and instead replicate them in the processor i.e., different stages of pipeline, buses, etc. They showed that if a feature vector was useful in detecting one target (seen variant), it is likely that a similar feature
detector in different positions in the pipeline can detect the evaded information (unseen variant). Replicated feature vectors also allows each patch of program to be represented in several
microarchitectural ways–making the trained model resilient to several evasions.
Feature replication greatly reduces the number of free parameters to be learned by
the model– enabling a simple HW design for fast classification.
Replicated features also allows simpler circuit for centralizing the signals as one signal can be captured by another signal down or up the pipeline through replicated feature detector design. 

\begin{figure}[!htbp]
    \centering
    \includegraphics[width=0.5\textwidth]{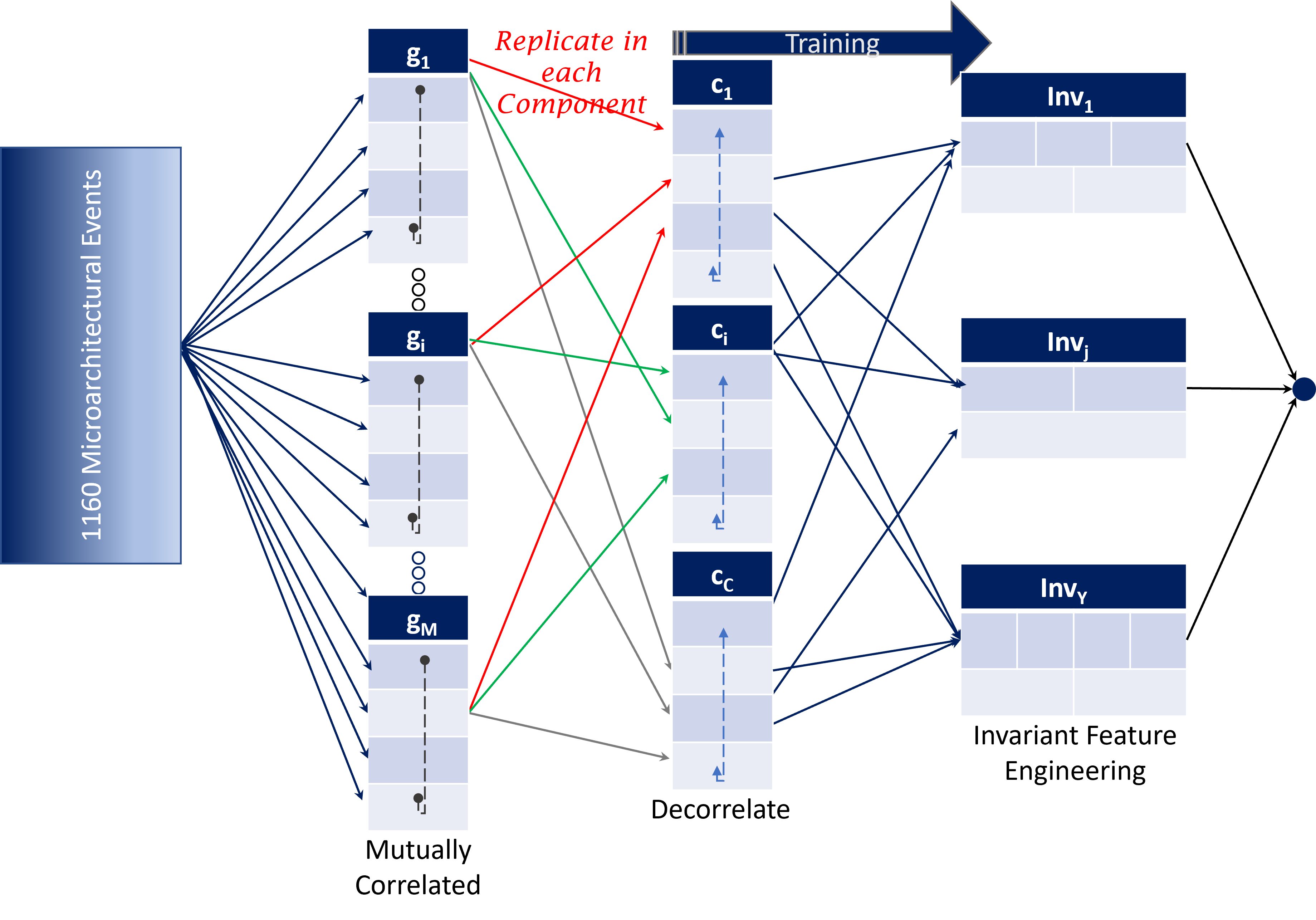}
    \caption{~\scheme{} Feature Selection (offline). Invariant Feature Engineering using GANs. }
    \label{fig:framework}
\end{figure}

\para{Perform the feature engineering offline}  Selecting replicated features and engineering new features is done  automatically through sub-sampling %
the hidden layers in the DNN Generator Network. 
To enable competitive performance with a shallow NN we apply the above steps of feature selection, replication, and feature engineering offline. 
This eliminates the need for hidden nodes and makes our detector more resilient to evasion. Figure~\ref{fig:framework} shows the overview of our framework. 

\begin{table}[!htbp]
\centering\small\resizebox{0.48\textwidth}{!}{
{\small\begin{tabular}{|c|c|}
\hline 
\#    &  Security HPCs   \\ \hline \hline
1 &  Squashed Bytes AND Bytes Read from WR Queue\\
2 &  Committed Maps AND rename.Undone \\
3 &  iew.Mem Order Violation AND dtlb.rdMisses \\
4 &  lsq.squashedStores AND lsq.forwLoads \\ 
5 & membus.trans\_dist::ReadSharedReq AND lsq.ignoredResponses\\
6 &  iq.SquashedNonSpecLD AND dcache.ReadReq\_mshr\_miss\_latency \\
7 & rename.serializingInsts   AND iew.ExecSquashedInsts \\

 \hline             
\end{tabular}%
}}
\caption{Subset of New Performance Counters for Security Engineered by \scheme{}.}
\label{tab:features}
\end{table}

\subsection{Hardware Overhead}
\begin{figure}[!htp]
\centering\includegraphics[width=0.5\textwidth]{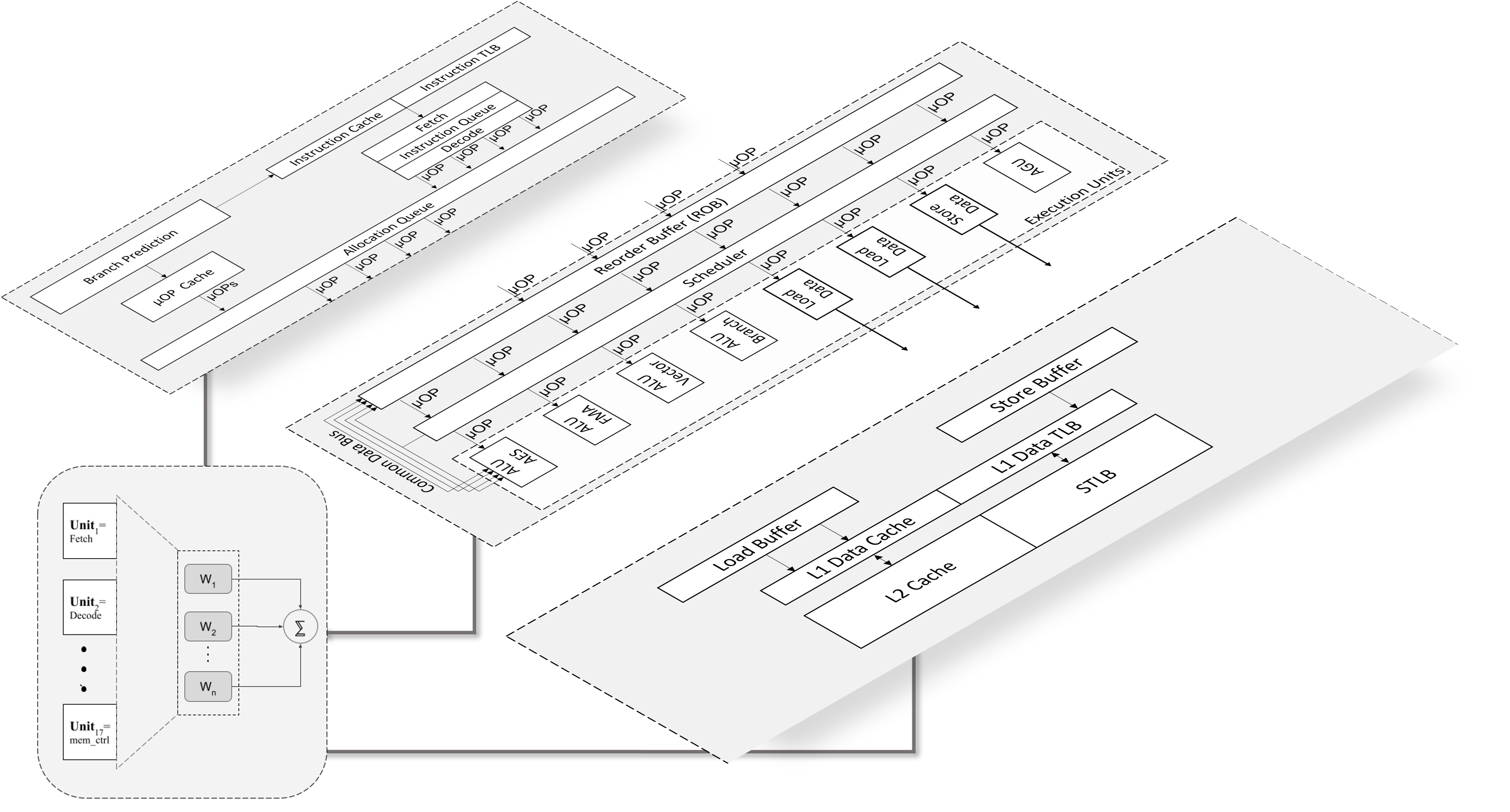}
  \caption{Hardware Detector}
  \label{fig:evax}
\end{figure}

We choose the perceptron neural network~\cite{HPCA2001Jimenez} because of its speed and small footprint. We also keep our hardware as similar to PerSpectron as possible to facilitate comparison.  Our training method can potentially improve the robustness of any ML based detection method. See section~\ref{dnn} for the result of training on AM-GAN samples for a DNN. A diagram of a Perceptron-based detector in the system is shown in Figure~\ref{fig:evax}.

We construct a quantized desired response $d(n)$ to a perceptron: $d(n)$ is 1 {if $x(n)$ belongs to malicious class} and 0 {if $x(n)$  belongs to the benign class}. The perceptron computes the weighted sum of the input patterns $x(n)$ 
comparing it to a threshold value.
If the sum exceeds the threshold, the output is +1; otherwise, it is 0. The central computation of perceptron-based inference, the dot-product computation, is performed using a modest circuit that adds partial products sequentially. 
Since \texttt{0} and \texttt{1} are the only possible input values, multiplication is unnecessary to compute the dot product. 
We only need to add a weight when the input bit is \texttt{1}. 
The sign bit of the result gives the prediction. 
The weights are static since \scheme{} uses offline training; they would only change based on a new security patch. 
The dot-product is not latency-sensitive, in contrast to the perceptron-based branch predictors~\cite{HPCA2001Jimenez,Exynosisca} that must employ expensive adder trees to compute the dot-product quickly.  
In our case, we need a single adder to add or subtract serially
each feature value. 
This will give us a result in a few hundred cycles in the worst case.%
Since the dot product can be computed sequentially, the hardware overhead of perceptron consists of three main components: 1) logic to find the dot product of the 145 weights and inputs, consisting mostly of a 9-bit adder and a register to accumulate the partial product, 2) storage for the weights and inputs, and 3) communication channels from the sources of the inputs to the storage for the inputs. 
{Weights are in the range of [-2,1] and there are 145 of them. So the range of the values is from -290 to +145, or 435 distinct values that can be stored in 9 bits. }
The first two components have almost negligible overhead. We estimate that the dot product computation logic would have a transistor count no more than 4,000, a tiny fraction of the logic used for a perceptron-based branch predictor. The communication of the 145 signals from the disparate parts of the processor core to the dot product computation also has a very low cost in terms of area, but the design complexity would be non-trivial. Still, the effort should be no worse than that required to support performance monitoring units ubiquitous in today's processors that take signals from all over the core to a central location.
Note that perceptron learning in hardware is practical for various applications, including branch prediction, prefetching, and replacement policies~\cite{intelISCA2019}. 
Recent microarchitectures from Oracle~\cite{SPARCT4}, AMD ({\em, e.g.} Bobcat, Jaguar, Piledriver, Zen, etc.), and Samsung~\cite{Mongoose,M3,Exynosisca} are documented to feature perceptron-based branch predictors.

\para{Weight \& Feature Updates. }
\scheme{} is capable of being updated via a vendor distributed patch. 
We anticipate newly emerging attacks in the future will require updates to neural weights and additions to the set of features being monitored. This is a process similar to microcode updates which is a secure feature available in current processors. 
With \textit{future proofing} in mind, we envision that all available hardware performance counters will be connected to our detector allowing us to update the neural weights and monitored feature set as new attacks emerge. Evax inherits all of these hardware advantages from PerSpectron.

\section{Methodology} %
\begin{table}[!t]

\centering
 \fontsize{8}{10}\selectfont
\begin{tabular}{|l|l|}
\hline

\textbf{Architecture}  &
X86 O3CPU 1 core Single Thread at 2.0GHz \\ \hline

\textbf{Core}  &
Tournament branch predictor,
16 RAS entries,\\ & 4096 BTB entries,
LQEntries=32, SQEntries=32,\\& ROBEntries=192, fetch/disp/issue/commit 8 wide\\
&
numPhysIntRegs=256,numPhysFloatRegs=256 \\ \hline

\textbf{L1 I-Cache}  &
32KB, 64B line, 4-way \\ \hline

\textbf{L1 D-Cache}  &
64KB, 64B line, 8-way \\ \hline

\textbf{L2 Shared Cache}  &
2MB bank, 64B line, 8-way, responsLatency=20\\
&
mshrs=20, tgtsPerMshr=12, writeBuffers=8  \\ 
&
tagLatency=20, dataLatency=20\\
\hline
\end{tabular}
\caption{Parameters of simulated architecture}\label{table:GEM5}
\end{table}

 \vspace*{-1.5mm}
\para{General Infrastructure.} We use the Gem5~\cite{gem5} cycle-level simulator to implement and evaluate the security and performance of our end-to-end defense.  We simulate O3CPU which is a detailed and highly configurable out of order CPU model (it does timing simulation for both CPU and memory which is essential for simulating microarchitectural side channel attacks). We simulate a full-System model  which include Operating System (Ubuntu 18.04 v. 4.8.13) and underlying architecture (X86 ISA). Our methodology should extend easily to other ISAs as well.
Table~\ref{table:GEM5} gives the parameters of the simulated architecture. 

\para{Attack Generation in gem5}. We use methods described in ~\cite{Gem5forcioli} to successfully implement our attacks in gem5 for security analysis.  
We also extended the supported instructions~\cite{gem5-avx} in gem5
to support new variants of microarchitectural attacks such as Medusa. 
Our modified gem5 uses an extensible DRAM simulator (Ramulator~\cite{kim2015ramulator}) which runs as part of a full-system gem5.
By design, gem5 and Ramulator
do not include the management of Rowhammer effects, so we
developed a dedicated memory corruption module and updated
some existing components similar to  ~\cite{Gem5forcioli} that stores an associative map to maintain the link between virtual and physical addresses. It determines the neighbors
of each row and establishes the affected ones, counts the number of activations in each row since the
last refresh, and affects one bit-flip threshold to each row. It establishes if one bit-flip occurs and modifies the affected
cells in consequence. These modifications make it possible to successfully run
existing Rowhammer attacks on an x86 platform associated with
classical DRAM. 
Additionally, we use the same disk image to compare
with real hardware, while keeping the same binaries and
libraries between the real world and the simulated world.
This makes testing attacks and ensuring simulation of security
features and side-channel sources (e.g., caches, DVFS) more
reproducible and easier than on a real platform.
We create 30 Simpoints per benchmark in full system mode. We warm up by 1M instructions 
and then start execution from the precise checkpoints running 1M 
instructions from each {(30M total)}. We simulate attacks to completion. 
Contrary to typical architectural studies, we generate many more, smaller simpoints of benign codes, since we need to train to detect short patterns
quickly. From gem5 we collect values of $1160$ microarchitectural counters, similar to prior work~\cite{PerSpectron}. 
For these events, we measure total number, cycles, rate, average, and distribution. 
For counters, we maintain a maximum seen value for each sampling simulation point. 
Statistics are normalized over the maximum value of the counter. 
\para{Infrastructure for ML \& HW Security Analysis. }
We have extended our framework to collect statistics once every 100,000, 10,000, 1000 and 100 instructions and sample all event counters for each program. We collect  performance counters from the Gem5 simulator to train our detector, and measure prediction accuracy and leakage information. We do not collect performance counters from SW because the available number of performance counters in SW is limited – Intel processors can monitor
four at a time. Many features correlated with malicious behavior are not available in software. GEM5 is not limited to the commit state and includes a rich feature set to monitor the speculative
execution state and we can compare with the state-of-art detector, which relies on sampling a large set of microarchitectural
features at once available in Gem5.
Using this tool, we collect multi-dimensional time-series traces of applications. 
We use the Keras library~\cite{keras} to implement and train our AM-GAN. 
We chose class Conditioned CGAN~\cite{cgan} as our reference GAN network. 
We used the FANN C library~\cite{FANNnissen03} to implement the final ~\scheme{} and PerSpectron models to work with Gem5.
 \begin{figure*}[!htbp]
    \centering
    \includegraphics[width=\textwidth]{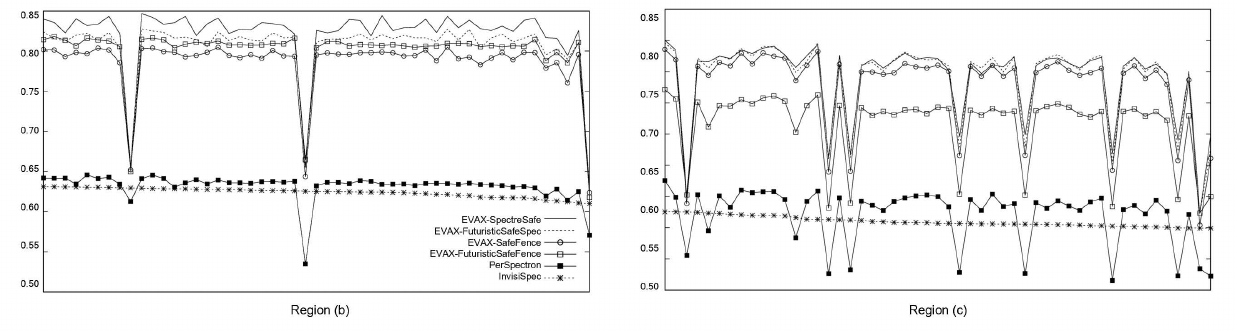}
  \caption{Performance (IPC) of an adaptive architecture with ~\scheme{} (vertical axis), compared to Perspectron and InvisiSpec.
  With~\scheme{}, we also examine increasingly conservative fencing schemes to anticipate future attacks.}

    \label{fig:IPCs}
\end{figure*}

\para{Infrastructure for Performance \& Security Analysis. }For performance analysis we have simulated two defenses implemented in GEM5 for the Spectre threat model. For Fencing we simulate a fence after every branch instruction. for InvisiSpec ~\cite{yan2018invisispec} we add support to track when instructions reach their visibility points, and issue load requests accordingly. We
add a SpecBuffer at each L1, and a SpecBuffer for LLC. 
We modified GEM5 to bypass these defenses during performance mode and switch back to them after each true flag.  We evaluate execution in secure mode for 10,000 instructions, 100,000 instructions and 1M instructions.

\para{Workload} We run individual SPEC CPU 2006 applications~\cite{spec2006}. 
The workloads include C compression programs modified to do most work in memory (rather than I/O), optimization scheduling, Ethernet network simulator, high-rank artificial intelligence programs, discrete event simulation, gene sequence protein analysis, the A* algorithm, and more. 
For transient attacks, we simulate Spectre-PHT~\cite{kocher2019spectre}, Spectre-BTB~\cite{spectre,canella2018systematic}, Spectre-RSB~\cite{koruyeh2018spectre}, Spectre-STL~\cite{SpectreV4}, Meltdown~\cite{meltdown}, three variants of Medusa~\cite{moghimi2020medusa}, Rowhammer~\cite{TRRespass,rowhammer}, SMotherSpectre~\cite{bhattacharyya2019smotherspectre}, 
LVI (non-SGX environment)~\cite{vanbulck2020lvi}, FallOut~\cite{canella2019fallout}, Branchscope~\cite{evtyushkin2018branchscope}, 
Microscope(non-SGX environment)~\cite{Microscope},  Leaky Buddies attack (CPU side)~\cite{dutta2020leaky}, RDRND~\cite{weber2021osiris}, FlushConflict~\cite{weber2021osiris} and DRAMA~\cite{TRRespass}. 
For cache attacks, we run Flush+Flush~\cite{GrussFlushFlush}, Flush+Reload~\cite{FlushReload2014Yarom},  Prime+Probe~\cite{osvik2006cache}. %

\para{Evasive Attacks. }
We generate 1.2M samples using our automated attack generation tools.  We use Transynther~\cite{moghimi2020medusa}, TRRespass~\cite{TRRespass} and Osiris~\cite{weber2021osiris} for automatic attack generation.
For our manual evasive attacks experiments, we applied techniques in malware~\cite{PaulKocher,kocher2019spectre} to  produce 29 attacks and 400,000 attack samples that can evade the state of the art hardware detector~\cite{PerSpectron}.
We produce a total of 257,066 attack samples (and 70,000 benign program samples to balance the data) using AM-GAN at each Cross Validation Fold. 
\para{Cross Validation Setting. }
We use K-fold cross validation for measuring mean accuracy on unseen attacks. At every fold, we remove all the samples belonging to one attack in the test set so that they are not used for model selection or AM-GAN training. 
For transient execution attacks, to ensure~\scheme{} detects the leakage and injection channel of the attacks, we also exclude the samples from recovery/transmission phase (check-pointed in our database) from the testing set of each fold. 
We use a set of fixed features. %
We don’t select new sets of features whenever we exclude a sample, but we retrain the weights at each fold.

\section{Security and Performance Analysis}
In this section we discuss the security provided by ~\scheme{}. 
We report the robustness of detector against new evasive technologies. We show the performance of detector for defense against AML attacks, after retraining the detector on samples generated by these tools.   We present detection accuracy for out-of-sample-attacks (zero-day setting). We then present the end-to-end solution with false positive and false negative results and with overall performance results compared to current always on mitigations.

\subsection{End-to-End Performance Results for Zero Leakage}\label{ete}
{In this section we present results for ML detectors trained on all the attacks in our dataset.}
To guarantee security, EVAX is tuned to have very high sensitivity. Therefore it detected all of the attacks in our training set before leakage. 
Figure~\ref{fig:IPCs} shows the performance of the adaptive architecture enabled by EVAX compared, to PerSpectron and InvisiSpec. We can see that the IPC for InvisiSpec is lower than  all the adaptive policies most of the time.  Perspectron performs slightly better than InvisiSpec but still at very low IPC. EVAX keeps the IPC  above 0.85 in most regions for mitigating Spectre (EVAX-SpectreSafe). IPC drops slightly for EVAX-FuturisticSafeFence which Fences all the loads. This architecture mitigates even the most dangerous class of microarchitectural attacks (e.g., LVI) with an overhead still significantly lower than prior mitigation techniques that are less aggressive.

\begin{figure}[!tp]
\centering
\includegraphics[width=0.35\textwidth]{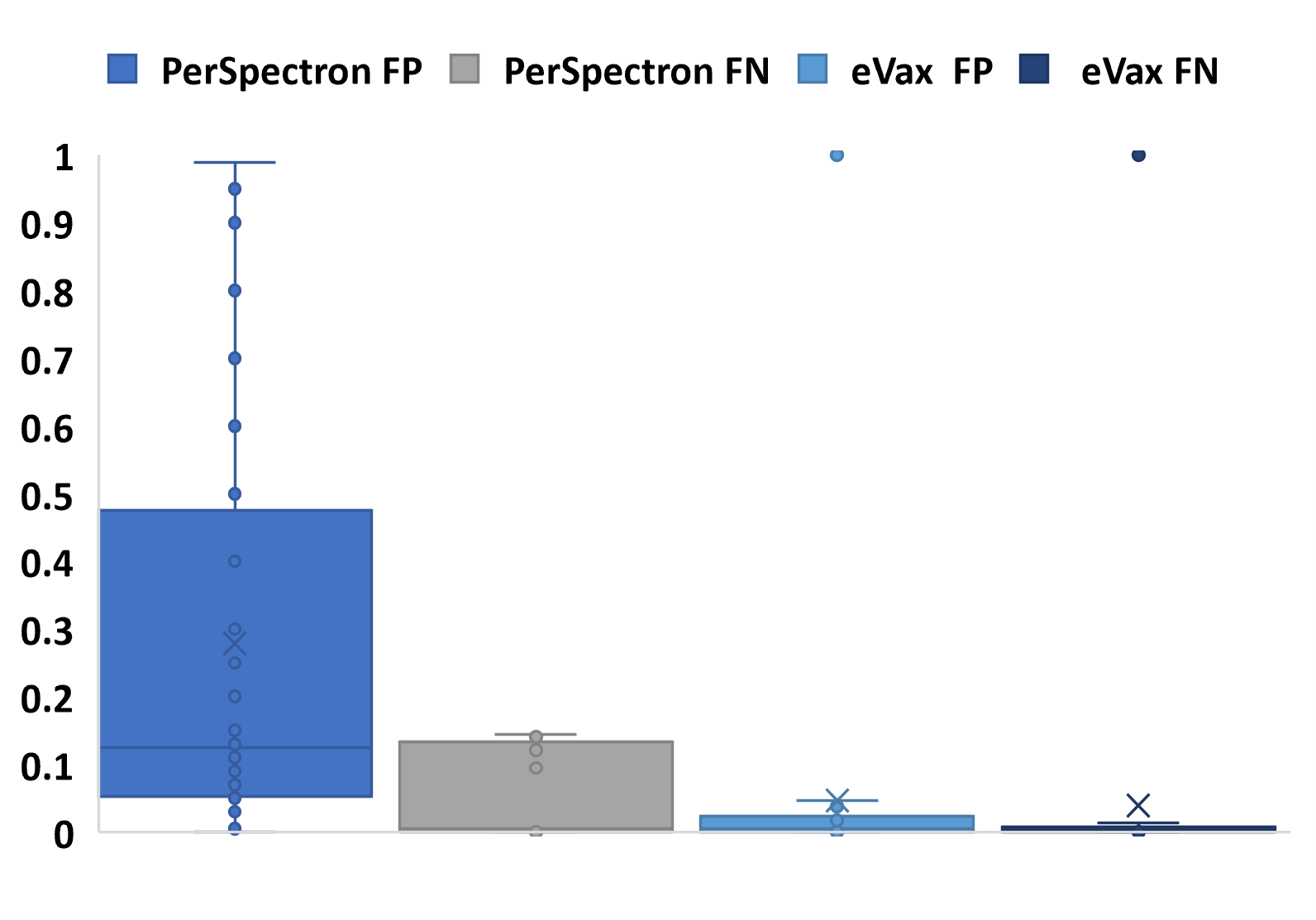}
\caption{False Positive (FP), False Negative (FN) Distribution 
} 
 \label{fig:fpfnbox}
 \end{figure}

\para{False Positives and Negatives (for Zero Leakage).} Figure~\ref{fig:fpfnbox} shows ~\scheme{} provides 85\% improvement in False Positives
(0.27 FN to 0.034) and 
72\% improvement in False Negatives
(0.11 FN to 0.03).
~\scheme{} has 0.034 FP in every 10K instruction, which
is 4 FPs in every 1M instruction. This is a practical False Positive for deployment.
This number is 0.0005 FP and 0.0001 FN for sampling every 100 instructions. 
This result is for a model that is also trained on a dataset of samples taken every 100 instructions. Our model's accuracy in detection of microarchitectural attacks  increases by higher sampling frequency which is expected since these attacks perform their anomalous behaviour during the transient window. 

  \begin{figure}[!hbp] 
\centering\includegraphics[width=0.5\textwidth]{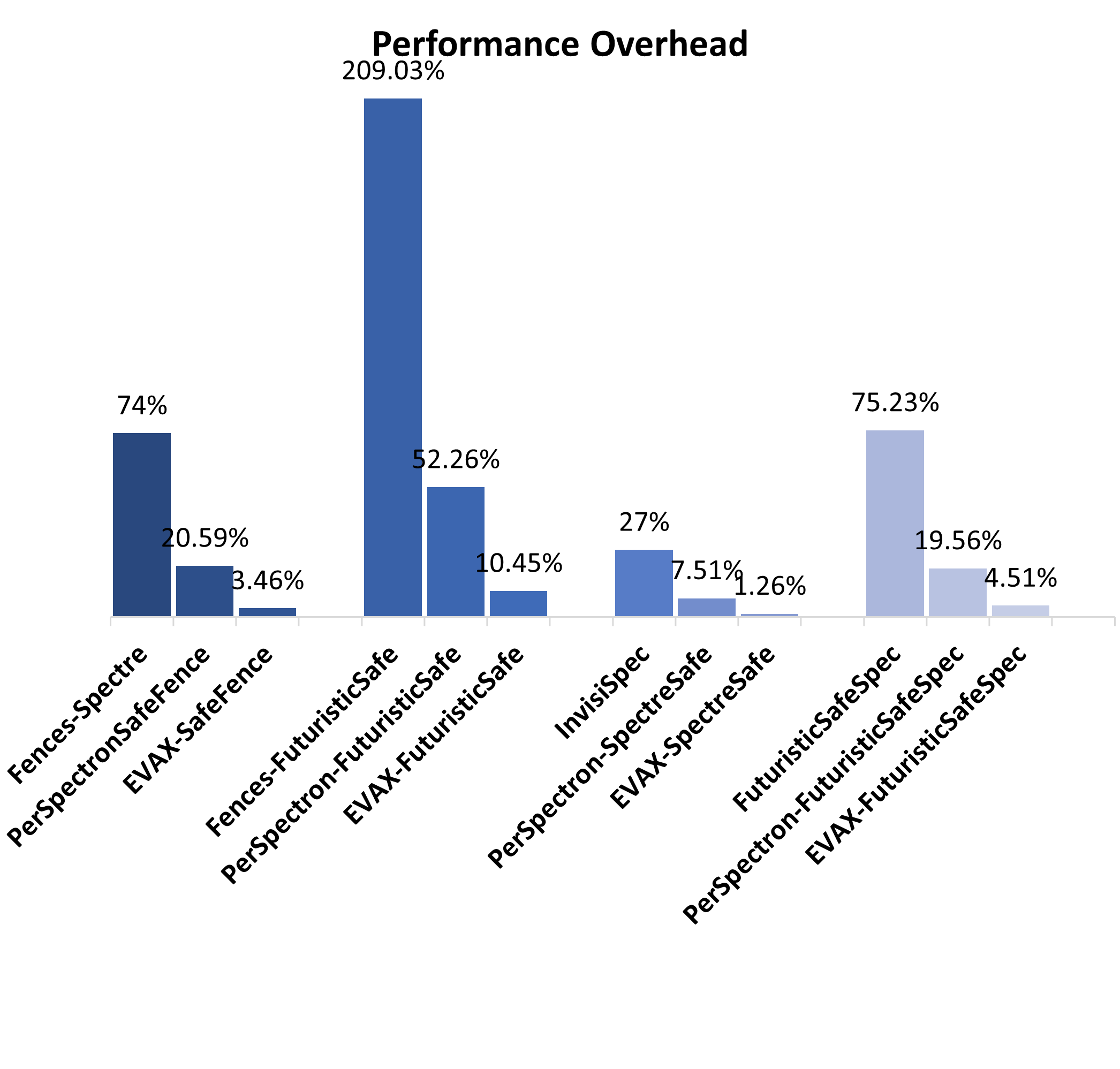}\vspace*{-1.5cm}\caption{End to end Defense Performance Comparison} 
\label{fig:perfMean}
\end{figure}

\para{Performance Overhead (for Full Security Coverage).}
For end to end results we turn on mitigation at every  true flag by our detector and we execute 1M instructions in secure mode to deactivate possible attacks. 
We only measure performance of benign program since performance of malicious programs is not a concern. 
We evaluate performance against hypothetical Futuristic Attacks Model that can exploit
any speculative load and Spectre Model that can exploit speculative loads follow an unresolved control-flow instruction ~\cite{yan2018invisispec}. 
Figure~\ref{fig:perfMean} shows an end to end defense performance comparison.  In the figure, Fences-FuturisticSafe simulates a fence before every load instruction, FuturisticSafeSpec is InvisiSpec working under Futuristic threat model.
 EVAX-FuturisticSafe enables Fences-FuturisticSafe and EVAX-SafeFence enables InvisiSpec. EVAX-SpectreSafe simulate a fence after every branch instruction.
Figure~\ref{fig:perfMean} shows that EVAX reduces Fencing overhead for Spectre mitigation from 74\% to 3.46\%  (95\% reduction). EVAX reduces InvisiSpec performance overhead for spectre mitigation from 27\% to 1.26\% (95\% reduction).
 We show that EVAX reduces PerSpectron's performance overhead by factor of 6 for Spectre Fencing and by a factor of 4 for Futuristic (including LVI) mitigation.
We show that EVAX reduces Fences overhead for Futuristic attack mitigation from 209\% to 10\%  (95\% reduction). 
EVAX reduces InvisiSpec overhead for Futuristic attack mitigation from 75\% to 4\% (94\% reduction). 

\subsection{Resiliency Against New Evasive Tech} 
\begin{figure}[!htp]
\centering\includegraphics[width=0.5\textwidth]{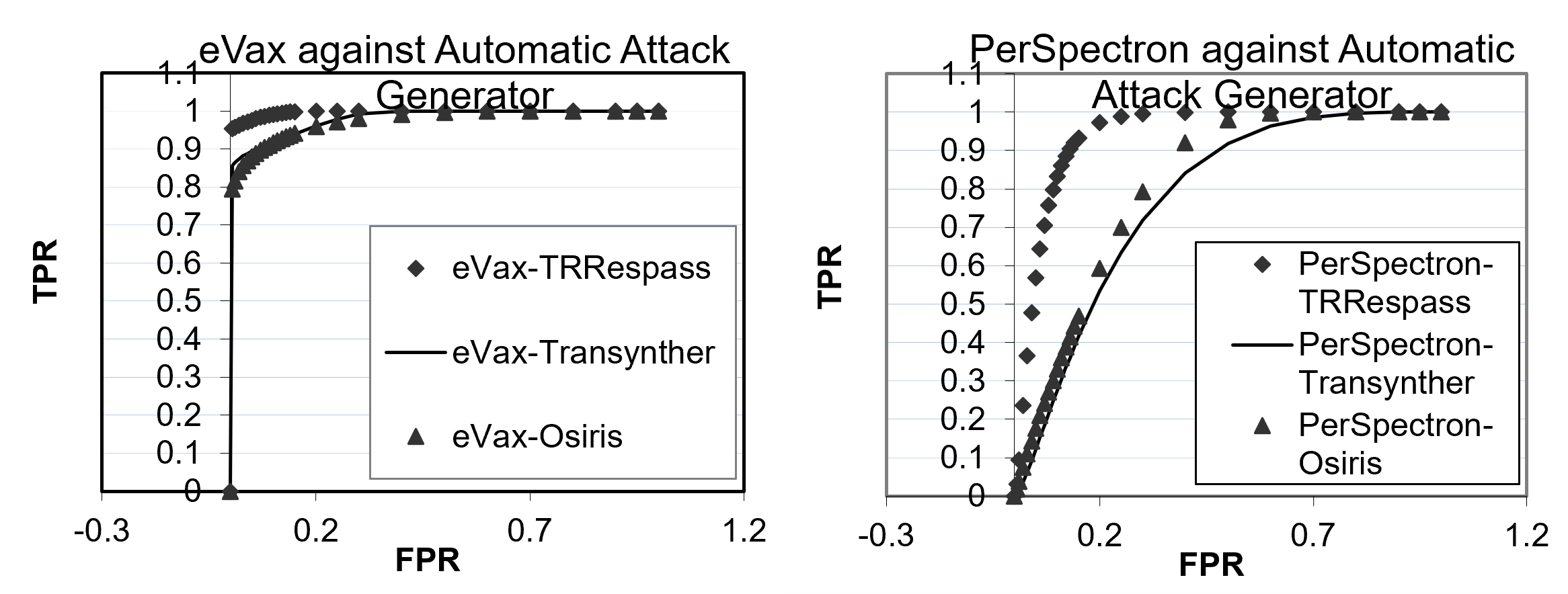}
\caption{Resiliency (ROC curve) Against 1.2M Evasive Attacks} 

 \label{fig:rocG}
\end{figure}

\begin{figure}[!hbp] 
\centering\includegraphics[width=0.3\textwidth]{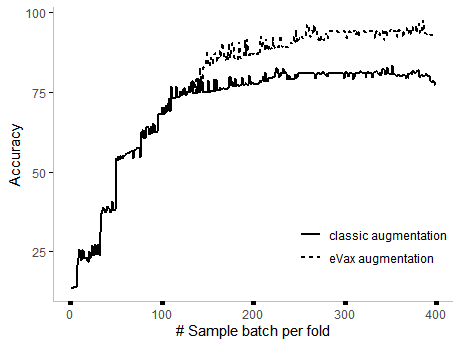}
\caption{ Filling the adversarial space. Increasing the accuracy on AML attacks. At 93\%, leakage is Zero. Thus all attempts to evade the detector disabled the attack. 
} 
\label{fig:accuracy}
\end{figure}

{In this experiment, we have set aside one third of the samples from our dataset for test and both detectors are trained on the remaining 2/3 of 
the attack samples in our dataset. None of the detectors are trained on the attacks generated by the automated tools~\cite{moghimi2020medusa, TRRespass, weber2021osiris}. } Figure~\ref{fig:rocG} shows PerSpectron under the attack by fuzzing tools that  break even the latest Intel processor.  These tools include Automated Rowhammer attacks (TRRespass), Automated fault-based attacks (Medusa), and Automated timing based attacks (Osiris).  We evaluate EVAX and PerSpectron on 1.2 Million attack samples generated by these powerful tools. We tune the EVAX output threshold to achieve detection resiliency as measured by the ROC curve.  %
The average Area Under the Curve
(AUC) of 0.797 for PerSpectron improves to 0.985 for EVAX {(23.5\% improvement).}
If we retrain the ML model on the same evasive tools, the accuracy on the adversarial attacks that evades the state-of-the-art plateaus at 78\% using Automated Attack Discovery Tools.   The accuracy improves to 93\% 
with EVAX (where leakage = Zero, FNs before leakage. ) {AM-GAN generated samples effectively cover the gap between PerSpectron classification boundary and the Microarchitectural attack leakage window (see Figure~\ref{fig:accuracy})}.

\begin{figure*}[!htbp]
    \centering
    \includegraphics[width=\textwidth]{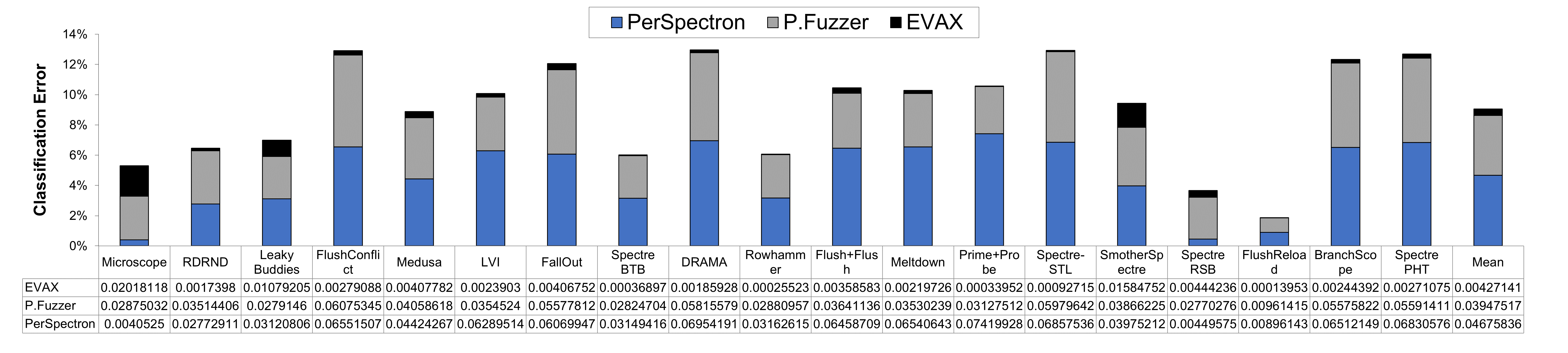}
\vspace{-0.6cm}\caption{
K-fold Cross Validation Setting}
    \label{fig:error}
\end{figure*}
\subsection{K-fold Cross Validation Setting}
 One of the main differences between attack samples generated by our AM-GAN and fuzzing approaches is that it tells the model that there are \textit{multiple correct answers}, {\em i.e.} multiple purturbation mechanisms or variations of a same attack, including those which haven't been seen (fuzzing is constrained by only a limited set of perturbations). This aligns with the objective of {\em seeking lower generalization error}, also known as the out-of-sample error or the risk. Generalization error is the conditional probability of correct prediction under an unknown state of parameters of the data generating mechanism,
Figure~\ref{fig:error} shows the classification error for EVAX, PerSpectron and PerSpectron hardened with sampling from the fuzzing tool (P.Fuzzer) in a zero day setting (K-fold cross validation). Evax drops the mean generalization error for PerSPectron, even when hardened by Fuzzing tools, by an order of magnitude.

{EVAX is exposed to not only many attack samples but also many generated \textit{benign program} samples through AM-GAN training of which represent comprehensive combination of normal utilization of  micro architectural component that exist in the safe programs. While AM-GAN differentiate different attack types to generate samples of each for training, our final classifier is binary. Everything that is not microarchitecturally benign is classified as a microarchitectural attack. So when the system learns a very large corpus of safe behaviour it improves its ability to identify many anomalous activities in regards to microarchitectural utilization which are mal-utilized by attacks.}
{Because AM-GAN generates a large  number (theoretically indefinite) samples of {\em benign} and {\em malicious} programs through the adversarial game setup, it eventually learns many of the constraints and functionality rules of the pipeline, and augments this knowledge to the detector.
For example, consider a feature   \textit{PendingQuiesceStallCycles} in fetch stage ($Q$) which is invariant to all kind of stalls down in the pipeline, e.g., flush, misses, and traps. $V=[f,m,t]$ shows the one hot representation of those counters. All of these stall signals propagate backwards to the fetch stage and are accumulated into the counter $Q$.}
{~\scheme{}'s benefit comes from the AM-GAN Generator which learns to generate all possible combinations of input vectors $V$ that trigger this {\em invariant feature} $Q$. 
 While the classical training (PerSpectron) learns the first two of three features that trigger $Q$. ~\scheme{} does draw the full connection between the third feature $t$ (traps) and $Q$ automatically through training on benign program samples generated by AM-GANs (new combination of seen behaviours into one that is not seen), allowing it to generalize to other fault-based attacks such as LVI (reverse of meltdown), Medusa and all evasive attacks generated by Transynther that PerSpectron fails to detect. . %
 We are able to see that the weight with value approximately $0$ 
 increases to value $2$ as AM-GAN trains on benign samples enabling the model to learn the structure of the pipeline.  We see this because Hyperplanes/perceptron are easy to understand and the first major step to cracking open a black box of deep learning.}
 
{Additionally, \scheme{} detects the RDRAND covert channel{~\cite{weber2021osiris}} with 95\% TPR detection when removed from the training set.
Weber et al.{~\cite{weber2021osiris}} showed that this attack is not easily detected nor prevented 
by any of the current software approaches{~\cite{payer2016hexpads}}{\cite{MASCAT}} {\cite{cacheBasedDetection2016Chiappetta}}{\cite{BlackHatFogh}}. 
Evax also detects FlushConflict in a Cross Validation setting, a microarchitectural kernel-level ASLR (KASLR) bypass that is not mitigated by any of the current hardware fixes \protect\cite{weber2021osiris}, with 97\% true positive rate (TPR) for ~\scheme{} and  63\% TPR for Perspectron. }
{This attack works even on the newest Intel Ice Lake and Comet Lake microarchitectures, even with all known mitigations in place. 
We also consider the Medusa attack \protect\cite{moghimi2020medusa}, a recent Meltdown-style attack that uses cache indexing, unaligned store-to-load forwarding, and shadow REP MOV in its different variations.
Note that Medusa evades PerSpectron by completing its meltdown basic block prior to detection.
\scheme{} detects all the meltdown-based steps of this attack with 98\% true-positive rate (TPR) detection for ~\scheme{} and 38\% TPR for Perspectron, demonstrating~\scheme{}’s robustness against advanced unseen meltdown-style and MDS attacks.}

{Interestingly, ~\scheme{}'s main contribution is the reverse of prior works. PerSpectron mapped high-dimension data to low-dimension $[f,m] \rightarrow V$ (classification), while AM-GAN's Generator maps low dimension data to high dimension $V\rightarrow [f,m,t]$ (generation).
This cross-validation experiment, when crafted carefully, can also test \scheme{}'s increased ability to detect zero-day attacks.
For example,
\scheme{} can generalize to detect DRAMA~\cite{TRRespass} with high accuracy (99\% TPR) even when it is excluded from the training set, while DRAMA evades PerSpectron.
Highly correlated features that are triggered during TRRespass detection include:   (1) \textit{selfRefreshEnergy}, (2) \textit{bytesPerActivate}, the number of accessed bytes per row activation in DRAM, (3) \textit{bytesReadWrQ}, the number of DRAM read requests serviced by the write queue. 
 None of these attacks were included in the ~\scheme{}'s training set at the fold they were being tested, showing ~\scheme{} could generalize to these attacks.}
{~\scheme{} cannot generalize to all zero-day attacks.
MicroScope, Leaky Buddies (CPU-side)  and  SMotherSpectre all evade detection (leak before detection) when not part of the train set. While not perfect, of course, Evax is a clear advance in the state of the art for detection of {\em zero-day attacks}.
However, if we retrain~\scheme{} with the samples held out from MicroScope, Leaky Buddies (CPU-side) and SMotherSpectre, we can detect them with 99.3\%, 99.0\% and 99.7\% accuracy indicating that there is a subset of shared features but not enough to allow {\em timely} detection in a cross validation setting.
This is why it is still important to provide the ability to update the training and feature set of the the detector when such feature-novel attacks appear. }

\subsection{Improving Other ML Detectors. } \label{dnn}
\begin{figure}[!htbp] 
\centering\includegraphics[width=0.45\textwidth]{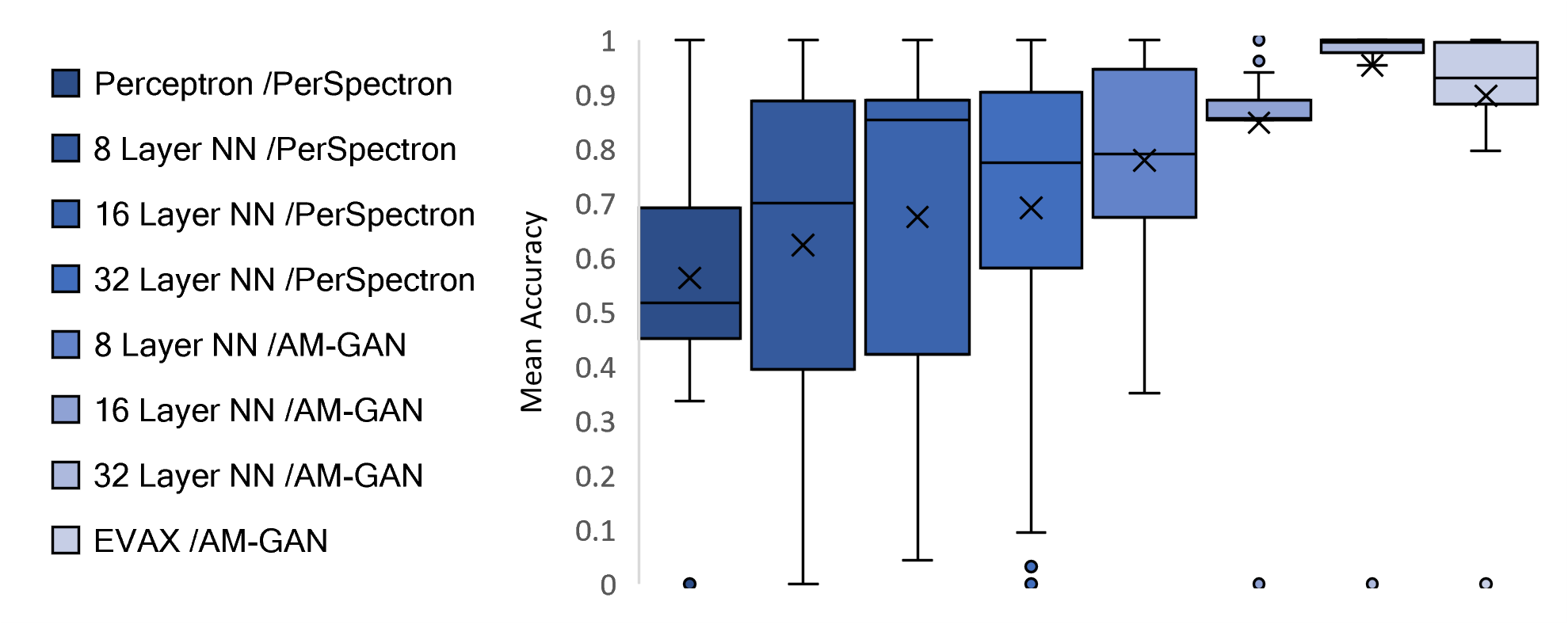}
\vspace{-0.2cm}\caption{  Improving other ML models with ~\scheme{}.
 } 
\label{fig:accuracyML}
\end{figure}

We focus on perceptrons because of their advantages for hardware implementation, fast detection and interpretability.  However, the techniques in this paper are applicable to a much
wider range of ML solutions for attack detection. 
Figure~\ref{fig:accuracyML} shows ~\scheme{}'s training  improves a 16-layer deep neural network's accuracy  significantly (range of 0.57-0.90 for traditional training to 0.95-0.99 for ~\scheme{} training).  Of particular significance is that our AM-GAN training enables a 16-layer neural network to outperform a 32-layer with it.
The effect of adding layers in traditional training, not only is not statistically significant  but also reduces the accuracy in some cases (Median of 85 for 16 Layer to 77 for 32 Layer). This is due to low quality training data.
One of the challenges in ML detector design for adaptive architectures is noise in the training data. For example, while the program phases can be check-pointed in code for reducing noise related to benign and malicious phases of programs, the syscall itself adds noise to the attack sample.  The CGAN
~\scheme{} framework is able to produce high quality, low-noise data for various types of attacks. This can remove the effect of noisy data and enable the capacity of deeper neural networks. One general take away is that increasing the complexity of neural networks without having a good set of training data can lead to statistically significant reduction in accuracy, as we see in Figure~\ref{fig:accuracyML}.  On the other hand, higher quality training data can significantly improve the capacity of shallower networks as the accuracy  of one layer NN with EVAX training is  significantly higher than even a 32 layer NN with traditional training for all workloads (Range 0.57-0.90 for 32 layer compared to 0.88-0.99 for EVAX).

\vspace{-0.35cm}\section{Conclusion}
EVAX proposes an end-to-end solution for runtime detection-based security against Microarchitectural attacks. {We showed that code transformation techniques used in the prior automatic attack generation works are not sufficient for training an ML detector. }This paper introduces a methodology for generating attack samples automatically in the microarchietctural layer rather than the code layer using GANs. The samples significantly increase the robustness of both simple and more complex microarchitectural attack detectors. EVAX also provides an automatic method to widen and linearize Deep NN by engineering features using trained GANs. Multiple catagories of attacks are then classified by a single hyperplane, which is fast and easily interpretable. We believe that linearization is the first major step to cracking open a black box of deep learning for systems.
The high sensitivity and specificity of EVAX and its fast, implementable HW and interpretable design, makes it practical for deployment, substantially lowering the high performance overhead of current state-of-the-art microarchitectural mitigations.
\section*{Acknowledgment}

We thank Daniel A. Jiménez for his comments during the drafting of this paper and Intel for a grant supporting this project.  

%

%
%
%
%
%
%
%
%
%
%
%
%
%
%
%
%
%
%
%
%
%
%

%
%

%
%
%
%
%
%
%
%
%
%
%
%
%
%
%
%
%
%
%
%
%

%
%
%
%
%
%
%
%
%
%
%
%
%
%
%
%
%
%
%
%
%
%
%

%
%
%
%
%
%

%

%
%
%
%
%
%
%

%

%
%
%
%
%
%
%
%
%
%
%
%
%

%
%
%

%


%
\end{document}